\documentclass[reqno]{amsart}

\usepackage{booktabs} 
\usepackage{IEEEtrantools}
\usepackage{amssymb,latexsym,amsfonts,amsmath}
\usepackage{mathrsfs}
\usepackage{graphicx}
\usepackage{dsfont}

\usepackage{tikz}
\usetikzlibrary{calc,shapes,arrows}

\usepackage{algorithm}
\usepackage{algorithmic}

\usepackage{xspace}

\usepackage[implicit=false]{hyperref}
\usepackage{listings}
\usepackage{lscape}

\usepackage[normalem]{ulem}

\newtheorem{theorem}{Theorem}[section]

\newtheorem{definition}[theorem]{Definition}

\newtheorem{remark}[theorem]{Remark}

\numberwithin{equation}{section}

\newcommand{\N}{{\mathbb{N}}}

\newcommand{\specialcell}[2][c]{%
	\begin{tabular}[#1]{@{}l@{}}#2\end{tabular}}

\begin{document}

\begin{abstract}
In this paper, we propose a software tool, called \textsf{AMYTISS}, implemented in C++/OpenCL, for designing correct-by-construction controllers for large-scale discrete-time stochastic systems.
This tool is employed to
(i) build finite Markov decision processes (MDPs) as finite abstractions of given original systems, and 
(ii) synthesize controllers for the constructed finite MDPs satisfying bounded-time high-level properties including safety, reachability and reach-avoid specifications. In \textsf{AMYTISS}, scalable parallel algorithms are designed such that they support the parallel execution within CPUs, GPUs and hardware accelerators (HWAs). Unlike all existing tools for stochastic systems, \textsf{AMYTISS} can utilize high-performance computing (HPC) platforms and cloud-computing services to mitigate the effects of the state-explosion problem, which is always present in analyzing large-scale stochastic systems. We benchmark \textsf{AMYTISS} against the most recent tools in the literature using several physical case studies including robot examples, room temperature and road traffic networks. We also apply our algorithms to a $3$-dimensional autonomous vehicle and $7$-dimensional nonlinear model of a BMW $320$i car by synthesizing an autonomous parking controller.
\end{abstract}

\title[\textsf{AMYTISS}: P\underline{a}rallelized Auto\underline{m}ated Controller S\underline{y}nthesis for Large-Scale S\underline{t}ochast\underline{i}c \underline{S}ystem\underline{s}]{\textsf{AMYTISS}: P\underline{a}rallelized Auto\underline{m}ated Controller S\underline{y}nthesis for Large-Scale S\underline{t}ochast\underline{i}c \underline{S}ystem\underline{s}}

\author{Abolfazl Lavaei$^{1,*}$}
\author{Mahmoud Khaled$^{2,*}$}
\author{Sadegh Soudjani$^3$}
\author{Majid Zamani$^{4,1}$}
\address{$^1$Department of Computer Science, Ludwig Maximilian University of Munich, Germany.}
\email{lavaei@lmu.de}
\address{$^2$Department of Electrical Engineering, Technical University of Munich, Germany.}
\email{khaled.mahmoud@tum.de}
\address{$^3$School of Computing, Newcastle University, UK.}
\email{sadegh.soudjani@ncl.ac.uk}
\address{$^4$Department of Computer Science, University of Colorado Boulder, USA.}
\email{majid.zamani@colorado.edu}
\maketitle

\section{Introduction}

\subsection{Motivations}
Large-scale stochastic systems are an important modeling framework to describe many real-life safety-critical systems such as power grids, traffic networks, self-driving cars, and many other applications. 
For this type of complex systems, automating the controller synthesis procedure to achieve high-level specifications, \emph{e.g.,} those expressed as linear temporal logic (LTL) formulae \cite{pnueli1977temporal}, is inherently very challenging mainly due to their computational complexity arising from uncountable sets of states and actions. To mitigate the encountered difficulty, finite abstractions, \emph{i.e.,} systems with finite state sets, are usually employed as replacements of original continuous-space systems in the controller synthesis procedure. 
More precisely, one can first abstract a given continuous-space system by a simpler one, \emph{e.g.,} a finite Markov decision process (MDP), and then perform analysis and synthesis over the abstract model (using algorithmic techniques from computer science \cite{baier2008principles}).
Finally, the results are carried back to the original system, while providing a guaranteed error bound \cite{lavaei2017compositional,lavaei2018ADHSJJ,lavaei2018ADHS,lavaei2017HSCC,lavaei2019NAHSJ,lavaei2018CDCJ,lavaei2019NAHS,lavaei2019LSS,lavaei2019HSCC_J,lavaei2019ECC,lavaei2019CDC,lavaei2019Thesis,MSSM19,HS_TAC19}.

Unfortunately, construction of finite MDPs for large-scale complex systems suffers severely from the so-called \emph{curse of dimensionality}: the computational complexity grows exponentially as the number of state variables increases. 
To alleviate this issue, one promising solution is to employ high-performance computing (HPC) platforms together with cloud-computing services to mitigate the state-explosion problem. In particular, HPC platforms have a large number of processing elements (PEs) and this significantly affects the time complexity when serial algorithms are parallelized \cite{jaja1992}.

\subsection{Contributions}
In this paper, we propose novel scalable parallel algorithms and efficient distributed data structures for first constructing finite MDPs of large-scale discrete-time stochastic systems. We then automate the computation of their correct-by-construction controllers given high-level specifications such as safety, reachability and reach-avoid. 
The main contributions and merits of this work are:
\begin{itemize}
	\item[(1)] 
	We propose a novel data-parallel algorithm for constructing finite MDPs from discrete-time stochastic systems and storing them in efficient distributed data containers. 
	The proposed algorithm handles large-scale systems.
	\item[(2)] 
	We propose a parallel algorithm for synthesizing discrete controllers using the constructed MDPs to satisfy safety, reachability, or reach-avoid specifications. More specifically, we introduce a parallel algorithm for the iterative computation of Bellman equation in standard dynamic programming~\cite{SSoudjani,esmaeil2013adaptive}.		    	
	\item [(3)]
	Unlike the existing tools in the literature, \textsf{AMYTISS} accepts bounded disturbances and natively supports both additive and multiplicative noises with different practical distributions including normal, uniform, exponential, and beta.
\end{itemize}		    
We apply the proposed implementations to real-world applications including robot examples, room temperature and road traffic networks, and autonomous vehicles. This extends the applicability of formal methods to some safety-critical real-world applications with high dimensions. The results show remarkable reductions in the memory usage and computation time outperforming all existing tools in the literature.

We provide \textsf{AMYTISS} as an \emph{open-source} tool.
After compilation, \textsf{AMYTISS} is loaded via \textsf{pFaces} \cite{pFaces} and launched for parallel execution within available parallel computing resources.
The source of \textsf{AMYTISS} and detailed instructions on its building and running can be found in:
\begin{center}
	\href{https://github.com/mkhaled87/pFaces-AMYTISS}{https://github.com/mkhaled87/pFaces-AMYTISS}
\end{center}

\subsection{Related Literature}
There exist several software tools on verification and synthesis of stochastic systems with different classes of models. \textsf{SReachTools}~\cite{vinod2019sreachtools} performs stochastic reachability analysis for linear, potentially time-varying, discrete-time stochastic systems. \textsf{ProbReach}~\cite{shmarov2015probreach} is a tool for verifying the probabilistic reachability for stochastic hybrid systems. \textsf{SReach}~\cite{wang2015sreach} solves
probabilistic bounded reachability problems for two classes of models: (i) nonlinear hybrid automata with parametric uncertainty, and (ii) probabilistic hybrid automata
with additional randomness for both transition probabilities and variable resets. \textsf{Modest Toolset}~\cite{hartmanns2014modest} performs modeling and analysis for hybrid, real-time, distributed and stochastic systems. Two competitions on tools for formal verification and policy synthesis of stochastic models are organized with reports in \cite{abate2018arch,abate2019arch}.

\textsf{FAUST}$^{\mathsf 2}$ \cite{FAUST15} generates formal abstractions for continuous-space discrete-time stochastic processes, and performs verification and synthesis for safety and reachability specifications. 
However, \textsf{FAUST}$^{\mathsf 2}$ is originally implemented in MATLAB and suffers from the curse of dimensionality due to its lack of scalability for large-scale models.
\textsf{StocHy} \cite{StocHy19} provides the quantitative analysis of discrete-time stochastic hybrid systems such that it constructs finite abstractions, and performs verification and synthesis for safety and reachability specifications.

\textsf{AMYTISS} differs from \textsf{FAUST}$^{\mathsf 2}$ and \textsf{StocHy} in two main directions. 
First, \textsf{AMYTISS} implements novel parallel algorithms and data structures targeting HPC platforms to reduce the undesirable effects of the state-explosion problem.  Accordingly, it is able to perform parallel execution in different heterogeneous computing platforms including CPUs, GPUs and HWAs. 
Whereas, \textsf{FAUST}$^{\mathsf 2}$ and \textsf{StocHy} can only run serially on one CPU, and consequently, it is limited to small systems. 
Additionally, \textsf{AMYTISS} can handle the abstraction construction and controller synthesis for two and a half player games (\emph{e.g.,} stochastic systems with bounded disturbances), whereas \textsf{FAUST}$^{\mathsf 2}$ and \textsf{StocHy} only handle one and a half player games (\emph{e.g.,} disturbance-free systems).

Unlike all existing tools, \textsf{AMYTISS} offers highly scalable, distributed execution of parallel algorithms utilizing all available processing elements (PEs) in any heterogeneous computing platform. 
To the best of our knowledge, \textsf{AMYTISS} is the only tool of its kind for continuous-space stochastic systems that is able to utilize all types of compute units (CUs), simultaneously.

We compare \textsf{AMYTISS} with \textsf{FAUST}$^{\mathsf 2}$ and \textsf{StocHy} in Table~\ref{comparison} in detail in terms of different technical aspects. 
Although there have been some efforts in \textsf{FAUST}$^{\mathsf 2}$ and \textsf{StocHy} for parallel implementations, these are not compatible with HPC platforms.
Specifically, \textsf{FAUST}$^{\mathsf 2}$ employs some parallelization techniques using parallel for-loops and sparse matrices inside Matlab, and \textsf{StocHy} uses \textsf{Armadillo}, a multi-threaded library for scientific computing.
However, these tools are not designed for the parallel computation on HPC platforms.
Consequently, they can only utilize CPUs and cannot run on GPUs or HWAs.
In comparison, \textsf{AMYTISS} is developed in OpenCL, a language specially designed for data-parallel tasks, and supports heterogeneous computing platforms combining CPUs, GPUs and HWAs.

	\begin{table}
	\small
	\centering				
	\caption{\small Comparison between \textsf{AMYTISS}, \textsf{FAUST}$^{\mathsf 2}$ and \textsf{StocHy} based on native features.}
	\vspace{2mm}
	\begin{tabular}{ p{2cm} | p{4.35cm} | p{4.1cm} | p{4.5cm} }
		\textbf{Aspect} & 
		\textbf{\textsf{FAUST}$^{\mathsf 2}$} & 
		\textbf{\textsf{StocHy}} & 
		\textbf{\textsf{AMYTISS}} \\ \hline
		\begin{tabular}[c]{@{}l@{}}Platform\end{tabular}      			 	  &  CPU   & CPU    &  All platforms  \\ \hline
		\begin{tabular}[c]{@{}l@{}}Algorithms\end{tabular}      		&  Serial on HPC	 & Serial on HPC   & Parallel on HPC \\ \hline
		\begin{tabular}[c]{@{}l@{}}Model\end{tabular} 		    &   Stochastic control systems: linear, bilinear  & Stochastic hybrid systems: linear, bilinear  & Stochastic control systems: nonlinear  \\ \hline
		\begin{tabular}[c]{@{}l@{}}Specification\end{tabular} & Safety, reachability & Safety, reachability & Safety, reachability, reach-avoid \\ \hline					
		\begin{tabular}[c]{@{}l@{}}Stochasticity\end{tabular} & Additive noise    	& Additive noise  & Additive \& multiplicative noises  \\ \hline		
		\begin{tabular}[c]{@{}l@{}}Distribution  \end{tabular} & Normal, user-defined& Normal, user-defined& Normal, uniform, exponential, beta, user-defined\\ \hline	
		\begin{tabular}[c]{@{}l@{}}Disturbance\end{tabular} & Not supported    	& Not supported  &  Supported  \\ \hline				
	\end{tabular}\vspace{0.3cm}
	\label{comparison}				
\end{table}

Note that \textsf{FAUST}$^{\mathsf 2}$ and \textsf{StocHy} do not natively support reach-avoid specifications in the sense that users can explicitly provide some avoid sets. Implementing this type of properties requires some modifications inside those tools. In addition, we do not make a comparison here with \textsf{SReachTools} since  it is mainly for stochastic reachability analysis of linear, potentially time-varying, discrete-time stochastic systems, while \textsf{AMYTISS} is not limited to reachability analysis and can handle nonlinear systems as well.

Note that we also provide a script in the tool repository\footnote[1]{\href{https://github.com/mkhaled87/pFaces-AMYTISS/blob/master/interface/exportPrismMDP.m}{https://github.com/mkhaled87/pFaces-AMYTISS/blob/master/interface/exportPrismMDP.m}} that converts the MDPs constructed by \textsf{AMYTISS} into \textsf{PRISM}-input-files~\cite{kwiatkowska2002prism}. In particular, \textsf{AMYTISS} can natively construct finite MDPs from continuous-space stochastic control systems. \textsf{PRISM} can then be employed to perform the controller synthesis for those classes of complex specifications that \textsf{AMYTISS} does not support.

\section{Notations and Preliminaries} 
We use the following notations throughout the paper. Sets of nonnegative and positive integers are denoted by $\mathbb N := \{0,1,2,\ldots\}$ and $\mathbb N_{\ge 1} := \{1,2,3,\ldots\}$, respectively. Notations $\mathbb R$, $ \mathbb R_{>0}$, and $\mathbb R_{\ge 0}$ denote, respectively, sets of real, positive and nonnegative real numbers. For any set $X$, we denote by $2^X$ the power set of $X$, \emph{i.e.,} the set of all subsets of $X$. We also denote by $\vert X \vert$ the cardinality of $X$. For an $n$-dimensional vector $x \in \mathbb R^{n}$, $x_i$, where $i\in\{1,\ldots,n\}$, denotes the $i$-th component of $x$.

Any $n$-dimensional hyper-rectangle (\emph{a.k.a.} hyper interval) is characterized by two corner vectors $x_{lb}, x_{ub} \in \mathbb{R}^n$ and we denote it by $[\![ x_{lb},x_{ub} ]\!] := [x_{lb,1},x_{ub,1}]\times[x_{lb,2},x_{ub,2}]\times\cdots\times[x_{lb,n},x_{ub,n}]$.
We denote by $\Vert x \Vert$ the infinity norm of $x$. Given $N$ vectors $x_i \in \mathbb R^{n_i}$, $n_i\in \mathbb N_{\ge 1}$, and $i\in\{1,\ldots,N\}$, we use $x = [x_1;\ldots;x_N]$ to denote the corresponding augmented vector of the dimension $\sum_i n_i$. 
Given a matrix $A$ in $\mathbb R^{n\times m}$, $A(:,b)$ denotes the $b$-th column of $A$, and $A(b,:)$ the $b$-th row of $A$.	

A probability space is a tuple $(\Omega,\mathcal F_{\Omega},\mathbb{P}_{\Omega})$, where $\Omega$ is the sample space,
$\mathcal F_{\Omega}$ is a $\sigma$-algebra on $\Omega$, which comprises subsets of $\Omega$ as events,
and $\mathbb{P}_{\Omega}$ is a probability measure that assigns probabilities to events.
A random variable $X$ is a measurable function $X:(\Omega,\mathcal F_{\Omega})\rightarrow (S_X,\mathcal F_X)$ inducing a probability measure on  its space $(S_X,\mathcal F_X)$ as $Prob\{A\} = \mathbb{P}_{\Omega}\{X^{-1}(A)\}$ for any $A\in \mathcal F_X$. 
We directly present the probability measure on $(S_X,\mathcal F_X)$ without explicitly mentioning the underlying probability space and the function $X$ itself.

A topological space $S$ is a Borel space if it is homeomorphic to a Borel subset of a Polish space (\emph{i.e.,} a separable and completely metrizable space). The Euclidean spaces $\mathbb R^n$, its Borel subsets endowed with a subspace topology, and hybrid spaces are examples of Borel spaces. A Borel $\sigma$-algebra is denoted by $\mathcal B(S)$, and any Borel space $S$ is assumed to be endowed with it. A map $f : S\rightarrow Y$ is measurable if it is Borel measurable.

\section{Discrete-Time Stochastic Control Systems}\label{sec:dt-SCS}
We formally introduce discrete-time stochastic control systems (dt-SCS) below.
\begin{definition}
	A discrete-time stochastic control system (dt-SCS) is a tuple
	\begin{equation}
	\label{eq:dt-SCS}
	\Sigma=\left(X,U,W,\varsigma,f\right)\!,
	\end{equation}
	where, 
	\begin{itemize}
		\item $X\subseteq \mathbb R^n$ is a Borel space as the state set and $(X, \mathcal B (X))$ is its measurable space;
		\item $U\subseteq \mathbb R^m$ is a Borel space as the input set;
		\item $W\subseteq \mathbb R^p$ is a Borel space as the disturbance set;
		\item $\varsigma$ is a sequence of independent and identically distributed (i.i.d.) random variables from a sample space $\Omega$ to a measurable set $\mathcal V_\varsigma$
		\begin{equation*}
		\varsigma:=\{\varsigma(k):\Omega\rightarrow \mathcal V_{\varsigma},\,\,k\in\N\};
		\end{equation*}
		\item $f:X\times U\times W\rightarrow X$ is a measurable function characterizing the state evolution of the system.
	\end{itemize}
\end{definition} 

The state evolution of $\Sigma$, for a given initial state $x(0)\in X$, an input sequence $\nu(\cdot):\mathbb N\rightarrow U$, and a disturbance sequence $w(\cdot):\mathbb N\rightarrow W$, is characterized by the difference equations
\begin{equation}\label{Eq_1a}
\Sigma:x(k+1)=f(x(k),\nu(k),w(k)) + \Upsilon(k),
\quad \quad k\in\mathbb N,
\end{equation}
where $\Upsilon(k) := \varsigma(k)$ with $\mathcal V_\varsigma = \mathbb R^n$ for the case of the additive noise, and $\Upsilon(k) := \varsigma(k)x(k)$ with $\mathcal V_\varsigma$ equals to the set of diagonal matrices of the dimension $n$ for the case of the multiplicative noise~\cite{li2005estimation}.
We keep the notation $\Sigma$ to indicate both cases and use respectively $\Sigma_\mathfrak a$ and $\Sigma_\mathfrak m$ when discussing these cases individually.

We should mention that our parallel algorithms are independent of the noise distribution.  For an easier presentation of the contribution, we present our algorithms and case studies based on normal distributions but our tool natively supports other practical distributions including uniform, exponential, and beta. In addition, we provide a subroutine in our software tool so that the user can still employ the parallel algorithms by providing the density function of the desired class of distributions.

We are interested in \emph{Markov policies} to control dt-SCS $\Sigma$ as defined below. 

\begin{definition}
	For the dt-SCS $\Sigma$ in \eqref{eq:dt-SCS}, a Markov policy is a sequence
	$\rho = (\rho_0,\rho_1,\rho_2,\ldots)$ of universally measurable stochastic kernels $\rho_n$ \cite{Bertsekas1996}, each defined on the input space $U$ given $X$ and such that for all $x_n\in X$, $\rho_n(U|x_n)=1$.
	The class of all such Markov policies is denoted by $\Pi_M$.
\end{definition}

\begin{remark}	
	Our synthesis is based on a $\max$-$\min$ optimization problem for two and a half player games by considering the disturbance and input of the system as players~\cite{kamgarpour2011discrete}. Particularly, we consider the disturbance affecting the system as an adversary and maximize the probability of satisfaction under the worst-case strategy of a rational adversary. Hence, we minimize the probability of satisfaction with respect to disturbances, and maximize it over control inputs.
\end{remark}

One may be interested in analyzing dt-SCSs without disturbances (cf. case studies). In this case, the tuple \eqref{eq:dt-SCS} reduces to 
\begin{align}\label{without_disturbance}
\Sigma=(X,U,\varsigma,f),
\end{align}
where $f:X\times U\rightarrow X$, and the equation \eqref{Eq_1a} can be re-written as
\begin{equation}\label{Eq_11a}
\Sigma:x(k+1)=f(x(k),\nu(k)) + \Upsilon(k),
\quad\quad k\in\mathbb N.
\end{equation}
Note that input models in this tool paper are given inside configuration text files. Systems are described by stochastic difference equations as~\eqref{Eq_1a}-\eqref{Eq_11a}, and the user should provide the right-hand-side of equations\footnote[2]{An example of such a configuration file is provided at:\\ \href{https://github.com/mkhaled87/pFaces-AMYTISS/blob/master/examples/ex_toy_safety/toy2d.cfg}{https://github.com/mkhaled87/pFaces-AMYTISS/blob/master/examples/ex-toy-safety/toy2d.cfg}}. In the next section, we formally define MDPs and discuss how to build finite MDPs from given dt-SCSs.

\section{Finite Markov Decision Processes (MDPs)} \label{subsec:MDP}

A dt-SCS $\Sigma$ in~\eqref{eq:dt-SCS} is \emph{equivalently} represented by the following MDP \cite[Proposition 7.6]{kallenberg1997foundations}:
\begin{equation}\notag
\Sigma=\left(X,U,W,T_{\mathsf x}\right)\!,	
\end{equation}
where the map $T_{\mathsf x}:\mathcal B(X)\times X\times U\times W\rightarrow[0,1]$, 
is a conditional stochastic kernel that assigns to any $x \in X$, $\nu\in U$, and $w\in W$\!, a probability measure $T_{\mathsf x}(\cdot | x,\nu, w)$
on the measurable space
$(X,\mathcal B(X))$
so that for any set $\mathcal{A} \in \mathcal B(X)$, 
$$\mathbb P (x(k+1)\in \mathcal{A}\,|\, x(k),\nu(k),w(k)) = \int_\mathcal{A} T_{\mathsf x} (\mathsf{d}x(k+1)|x(k),\nu(k),w(k)).$$
For given input $\nu(\cdot),$ and disturbance $w(\cdot),$  the stochastic kernel $T_{\mathsf x}$ captures the evolution of the state of $\Sigma$ and can be uniquely determined by the pair $(\varsigma,f)$ from \eqref{eq:dt-SCS}. In other words, $T_{\mathsf x}$ contains the information of the function $f$ and the distribution of noise $\varsigma(\cdot)$ in the dynamical representation.

The alternative representation as the MDP is utilized in \cite{SAM15} to approximate a dt-SCS $\Sigma$ with a \emph{finite} MDP $\widehat\Sigma$ using an abstraction algorithm. This algorithm first constructs a finite partition of the state set $X = \cup_i \mathsf X_i$, the input set $U = \cup_i \mathsf U_i$, and the disturbance set $W = \cup_i \mathsf W_i$.
Then representative points $\bar x_i\in \mathsf X_i$, $\bar \nu_i\in \mathsf U_i$, and $\bar w_i\in \mathsf W_i$ are selected as abstract states, inputs, and disturbances.
The transition probability matrix for the finite MDP $\widehat\Sigma$ is also computed as
\begin{equation}
\label{eq:trans_prob}
\hat T_{\mathsf x} (x'|x,\nu,w) 
= T_{\mathsf x} (\Xi(x')|x,\nu,w),
\end{equation}
$\forall x, x'\in \hat X, \forall\nu\in \hat U, \forall w\in \hat W$, where the map $\Xi:X\rightarrow 2^X$ assigns to any $x\in X$, the corresponding partition element it belongs to, \emph{i.e.,} $\Xi(x) = \mathsf X_i$ if $x\in \mathsf X_i$. Since $\hat X$, $\hat U$ and $\hat W$ are finite sets, $\hat T_{\mathsf x}$ is a static map.
It can be represented with a matrix and we refer to it, from now on, as the transition probability matrix.

Given a dt-SCS $\Sigma=\left(X,U,W,\varsigma,f\right)$,
the finite MDP $\widehat\Sigma$ can be represented as a finite dt-SCS
\begin{equation}
\label{eq:abs_tuple}
\widehat\Sigma =(\hat X, \hat U,\hat W, \varsigma,\hat f),
\end{equation}
where $\hat f:\hat X\times\hat U\times\hat W \rightarrow\hat X$ is defined as
\begin{equation*}
\hat f(\hat{x},\hat{\nu},\hat{w}) = \Pi_x(f(\hat{x},\hat{\nu},\hat{w})),	
\end{equation*}
and $\Pi_x:X\rightarrow \hat X$ is a map that assigns to any $x\in X$, the representative point $\bar x\in\hat X$ of the corresponding partition set containing $x$. The Map $\Pi_x$ satisfies the inequality
\begin{equation}
\label{eq:Pi_delta}
\Vert \Pi_x(x)-x\Vert \leq \delta,\quad \forall x\in X,
\end{equation}
where $\delta:=\sup\{\|x-x'\|,\,\, x,x'\in \mathsf X_i,\,i=1,2,\ldots,n_x\}$ is the \emph{state} discretization parameter.
The initial state of $\widehat\Sigma$ is also selected according to $\hat x_0 := \Pi_x(x_0)$ with $x_0$ being the initial state of $\Sigma$. 

For a given logic specification $\varphi$ and accuracy level $\epsilon$, the discretization parameter $\delta$ can be selected a priori such that
\begin{equation}
\label{eq:metric_lit}
|\mathbb P(\Sigma\vDash\varphi) - \mathbb P(\widehat\Sigma\vDash\varphi)|\le \epsilon,
\end{equation}
where $\epsilon$ depends on the horizon of formula $\varphi$, the Lipschitz constant of the stochastic kernel, and $\delta$ (cf. \cite[Theorem 9]{SAM15}).

In the next sections, we propose novel parallel algorithms for the construction of finite MDPs and the synthesis of their controllers.

\section{Parallel Construction of Finite MDPs}\label{Parallel_Finite_MDPs}
In this section, we propose an approach to efficiently compute the transition probability matrix $\hat T_{\mathsf x}$ of the finite MDP $\widehat \Sigma$, 
which is essential for any controller synthesis procedure, as we discuss later in Section \ref{Parallel_Controllers_Synthesis}.

Algorithm \ref{alg:serial_T_computation} presents the traditional serial algorithm for computing $\hat T_{\mathsf x}$.
Note that if there are no disturbances in the given dynamics as discussed in \eqref{without_disturbance}, one can still employ 
Algorithm \ref{alg:serial_T_computation} to compute the transition probability matrix but without step \ref{step:sa_for_w}.\vspace{0.2cm}

\begin{algorithm}[ht]
	\caption{\small Traditional \emph{serial} algorithm for computing $\hat T_{\mathsf x}$}
	\label{alg:serial_T_computation}
	\begin{center}
		\begin{algorithmic}[1]
			\REQUIRE $\hat X, \hat U, \hat W$, and a noise covariance matrix $\mathsf{\Sigma} \in \mathbb R^{n\times n}$
			\ENSURE Transition probability matrix $\hat T_{\mathsf x}$ with the dimension of $(n_x\times n_\nu\times n_w,n_x)$ 
			\STATE \label{step:sa_for_x} {\bf for all} $\bar x_i \in \hat X, \text{s.t. } i \in \{1,...,n_x\},$ {\bf do} 
			\STATE \label{step:sa_for_u} \quad {\bf for all} $\bar\nu_j \in \hat U, \text{s.t. } j \in \{1,...,n_{\nu}\},$ {\bf do}	
			\STATE \label{step:sa_for_w} \quad \quad {\bf for all}  $\bar w_k \in \hat W, \text{s.t. } k \in \{1,...,n_w\},$ {\bf do} 
			\STATE \quad \quad \quad \label{step:sa_compute_mu} Compute mean $\mu$ as $\mu = f(\bar x_i,\bar \nu_j,\bar w_k, 0)$
			\STATE \label{step:sa_for_x_prime} \quad \quad \quad {\bf for all} $\bar x'_l \in \hat X, \text{s.t. } l \in \{1,...,n_x\},$ {\bf do} 								
			\begin{align}\notag
			\hat T_{\mathsf x}(\bar x'_l \vert \bar x_i, \bar {\nu}_j, \bar w_k) &:= \int_{\Xi(x')} \text{PDF}(\mathsf{d}x|\,\mu, \mathsf{\Sigma}),
			\end{align}
			\quad \quad \quad where PDF here is the probability density function of the normal distribution.
			\STATE \quad \quad \quad {\bf end}				
			\STATE \quad \quad {\bf end}
			\STATE \quad {\bf end}	
			\STATE {\bf end}	
		\end{algorithmic}
	\end{center}
\end{algorithm}\vspace{0.2cm}

In Subsections \ref{ssec:MDP_improve_parallel} and \ref{ssec:MDP_improve_cutting_threshold}, we address improvements of Algorithm \ref{alg:serial_T_computation}.
Each subsection targets one inefficient aspect of Algorithm \ref{alg:serial_T_computation} and discusses how to improve it.
In Subsection \ref{ssec:MDP_parallel_algorithem}, we combine the proposed improvements and introduce a parallel algorithm for constructing $\hat T_{\mathsf x}$. 

\subsection{Data-Parallel Threads for Computing $\hat T_{\mathsf x}$}
\label{ssec:MDP_improve_parallel}
The inner steps inside the nested for-loops \ref{step:sa_for_x}, \ref{step:sa_for_u}, and \ref{step:sa_for_w} in Algorithm \ref{alg:serial_T_computation} are computationally independent. More specifically, the computations of mean $\mu = f(\bar x_i,\bar \nu_j,\bar w_k, 0)$, $\text{PDF}(x\,|\,\mu, \mathsf{\Sigma})$, where PDF stands for probability density functions and $\mathsf{\Sigma}$ is a noise covariance matrix, and of $\hat T_{\mathsf x}$ all do not share data from one inner-loop to another. Hence, this is an embarrassingly data-parallel section of the algorithm. \textsf{pFaces} \cite{pFaces} can be utilized to launch necessary number of parallel threads on the employed hardware configuration (HWC) to improve the computation time of the algorithm. Each thread will eventually compute and store, independently, its corresponding values within $\hat T_{\mathsf x}$.

\subsection{Less Memory for Post States in $\hat T_{\mathsf x}$}\label{ssec:MDP_improve_cutting_threshold}

$\hat T_{\mathsf x}$ is a matrix with the dimension of $(n_x\times n_\nu\times n_w,n_x)$. The number of columns is $n_x$ as we need to compute and store the probability for each reachable partition element $\Xi(x'_l)$, corresponding to the representing post state $x'_l$. Here, we consider the Gaussian PDFs for the sake of a simpler presentation. For simplicity, we now focus on the computation of tuple $(\bar x_i, \bar \nu_j, \bar w_k)$. In many cases, when the PDF is decaying fast, only partition elements near $\mu$ have high probabilities of being reached, starting from $\bar x_i$ and applying an input $\bar \nu_j$.

\begin{figure}[ht]
	\begin{center}
		\includegraphics[width=11cm]{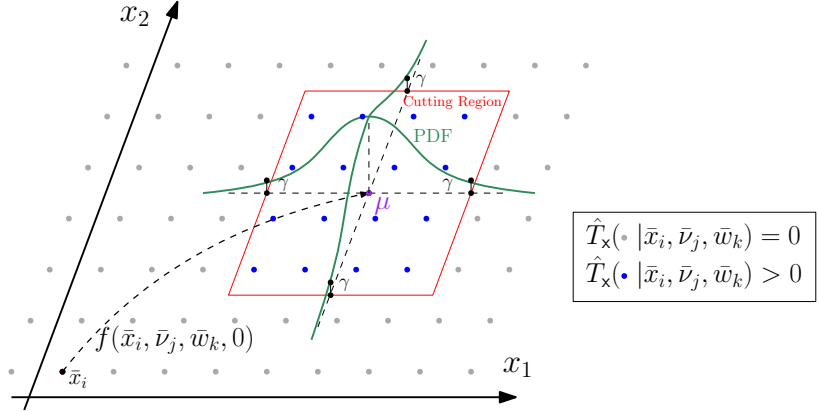}
		\caption{
			\small A 2-dimensional visualization of the cutting probability region (shown in red) with a cutting threshold of $\gamma$. 
			The cutting region encloses representative post states (blue dots) that have non-zero probabilities in $\hat T_{\mathsf x}$. 
			Other post states outside of the cutting region are considered to have zero probabilities in $\hat T_{\mathsf x}$.
		}
		\label{fig_cutting_region}
	\end{center}
\end{figure}

We set a cutting probability threshold $\gamma \in [0,1]$ to control how many partition elements around $\mu$ should be stored.
For a given mean value $\mu$, a covariance matrix $\mathsf{\Sigma}$ and a cutting probability threshold $\gamma$, $x \in X$ is called a PDF cutting point if $\gamma = \text{PDF}(x \vert \mu, \mathsf{\Sigma})$.
Since  Gaussian PDFs are symmetric, by repeating this cutting process dimension-wise, we end up with a set of points forming a hyper-rectangle in $X$, which we call it the cutting region and denote it by $\hat{X}^{\mathsf{\Sigma}}_{\gamma}$.
This is visualized in Figure \ref{fig_cutting_region} for a 2-dimensional system. Any partition element $\Xi(x'_l)$ with $x'_l$ outside the cutting region is considered to have zero probability of being reached.
Such approximation allows controlling the sparsity of the columns of $\hat T_{\mathsf x}$.
The closer the value of $\gamma$ to zero, the more accurate $\hat T_{\mathsf x}$ in representing transitions of $\widehat\Sigma$.
On the other hand, the closer the value of $\gamma$ to one, less post state values need to be stored as columns in $\hat T_{\mathsf x}$.
The number of probabilities to be stored for each $(\bar x_i, \bar \nu_j, \bar w_k)$ is then $\vert \hat{X}^{\mathsf{\Sigma}}_{\gamma} \vert$.
Figure \ref{fig_cutting_region} visualizes how the proposed $\gamma$ can help controlling the required memory for storing the transitions in $\hat T_{\mathsf x}$.

Note that since $\mathsf{\Sigma}$ is fixed prior to running the algorithm, number of columns needed for a fixed $\gamma$ can be identified before launching the computation. We can then accurately allocate a uniform fixed number of memory locations for any tuple $(\bar x_i,  \bar \nu_j, \bar w_k)$ in $\hat T_{\mathsf x}$.
Hence, there is no need for a dynamic sparse matrix data structure and  $\hat T_{\mathsf x}$ is now a matrix with a dimension of $(n_x\times n_\nu\times n_w,\vert \hat{X}^{\mathsf{\Sigma}}_{\gamma} \vert)$.

\begin{remark}
	\label{remark:constructing_the_cutting_region}
	Construction of $\hat{X}^{\mathsf{\Sigma}}_{\gamma}$ is practically a simple process.
	We start by solving the equation $\text{PDF}(x^*\,|\,0, \mathsf{\Sigma}) = \gamma$ for $x^* \in \mathbb{R}^n_{>0}$ and computing the zero-mean cutting points at each dimension. 
	Now, since the PDF is symmetric, one obtains 
	\begin{equation}
	\centering
	\nonumber
	\hat{X}^{\mathsf{\Sigma}}_{\gamma} = \{ \bar x \in \hat{X} \vert \bar x \in [\![\mu-x^*,\mu+x^*]\!] \}.
	\end{equation}
\end{remark}			

\begin{remark}
	The reduction in memory usage discussed in this subsection is tailored to Gaussian distributions for the sake of a better presentation of the idea. Users interested in adding additional distributions to \textsf{AMYTISS} have the option of providing a subroutine that describes how other distributions should behave in terms of required memory and with respect to the cutting threshold $\gamma$.
\end{remark}							

\subsection{A Parallel Algorithm for Constructing Finite MDP $\widehat{\Sigma}$}
\label{ssec:MDP_parallel_algorithem}

We present a novel parallel algorithm (Algorithm \ref{alg:parallell_T_computation}) to efficiently construct and store $\hat T_{\mathsf x}$ as a successor to Algorithm \ref{alg:serial_T_computation}.
We employ the discussed enhancements in Subsections \ref{ssec:MDP_improve_parallel} and \ref{ssec:MDP_improve_cutting_threshold} within the proposed algorithm.			
We do not parallelize the for-loop in Algorithm 2, Step 2, to avoid excessive parallelism (\emph{i.e.,} we parallelize loops only over $X$ and $U$, but not over $W$). Note that, practically, for large-scale systems, $\vert \hat X \times \hat U \vert$ can reach up to billions. We are interested in the number of parallel threads that can be scheduled reasonably by available HW computing units.\vspace{0.2cm}

\begin{algorithm}[ht]
	\caption{\small Proposed \emph{parallel} algorithm for computing $\hat T_{\mathsf x}$}
	\label{alg:parallell_T_computation}
	\begin{center}
		\begin{algorithmic}[1]
			\REQUIRE $\hat X, \hat U, \hat W, \gamma$, and a noise covariance matrix $\mathsf{\Sigma} \in \mathbb R^{n\times n}$
			\ENSURE Transition probability matrix $\hat T_{\mathsf x}$ with the dimension of $(n_x\times n_\nu\times n_w,\vert \hat{X}^{\mathsf{\Sigma}, W}_{\gamma} \vert)$ 
			\STATE {\bf for all} $(\bar x, \bar \nu) \in \hat X \times \hat U$ {\bf \emph{in parallel} do}			
			\STATE  \label{step:pa_for_w} \quad {\bf for all} $\bar w \in \hat W$ {\bf do}
			\STATE  \quad \quad Set $\mu = f(\bar x, \bar \nu, \bar w)$
			\STATE  \quad \quad Construct $\hat{X}^{\mathsf{\Sigma}}_{\gamma}$ as described in Remark \ref{remark:constructing_the_cutting_region}
			\STATE  \quad \quad {\bf for all} $x^* \in \hat{X}^{\mathsf{\Sigma}}_{\gamma}$ {\bf do}
			\STATE  \quad \quad \quad Set $\hat T_{\mathsf x} (x^*|\bar x,\bar \nu,\bar w) := \int_{\Xi(x^*)} \text{PDF}(\mathsf{d}x \vert \mu, \mathsf{\Sigma})$									
			\STATE  \quad \quad {\bf end} 						
			\STATE  \quad {\bf end} 
			\STATE {\bf end} 
		\end{algorithmic}
	\end{center}
\end{algorithm}

\section{Parallel Synthesis of Controllers}\label{Parallel_Controllers_Synthesis}
In this section, we employ dynamic programming to synthesize controllers for constructed finite MDPs $\widehat\Sigma$ satisfying safety, reachability, and reach-avoid properties \cite{SSoudjani,esmaeil2013adaptive}. 
We first present the traditional serial algorithm in Algorithm \ref{alg:serial_synthesis}. 
Note that if there are no disturbances in the given dynamics, Steps \ref{step:sa_synthesis_min_wrt_W} and~\ref{step_1:sa_synthesis_min_wrt_W} of Algorithm \ref{alg:serial_synthesis} are to be excluded.

\begin{algorithm}[ht]
	\caption{\small Traditional \emph{serial} algorithm for controller synthesis satisfying safety, reachability and reach-avoid specifications}
	\label{alg:serial_synthesis}
	\begin{center}
		\begin{algorithmic}[1]
			\REQUIRE $\hat X$, $\hat U$, $\hat W$, bounded time horizon $T_d$, $specs \in \{ Safety, Reachability,$ $Reach$-$Avoid \}$, target set $\mathcal{T}$ (in case $specs = Reachability, Reach$-$Avoid$), and avoid set $\mathcal{A}$ (in case $specs = Reach$-$Avoid$).
			\ENSURE Optimal satisfaction probability $V_v$ at time step $T_d = 1$, and optimal policy $\nu^\star$ corresponding to the optimal satisfaction probability.
			
			\STATE Compute $\hat T_{\mathsf x}$ as presented in Algorithm \ref{alg:serial_T_computation}.
			\STATE {\bf if} $specs == Safety$ {\bf do} 
			\STATE \quad \quad Set value function $V_v := ones(n_x,T_d+1)$
			\STATE {\bf else}
			\STATE \quad \quad Compute a transition probability matrix $\hat T_{0\mathsf x}$ from $\hat X \backslash (\mathcal{T} \cup \mathcal{A})$ to $\mathcal{T}$
			\STATE \quad \quad Set $\hat T_{\mathsf x}$ to zero for any post-state in $(\mathcal{T} \cup \mathcal{A})$.
			\STATE \quad \quad Set value function $V_v := zeros(n_x,T_d+1)$
			\STATE {\bf end}
			
			\STATE \label{step:sa_synthesis_for_time} {\bf for}  $k = T_d:-1:1$ (\emph{backward in time}) {\bf do} 	
			\STATE  \quad \quad {\bf if} $specs == Safety$ {\bf do} 
			\STATE \label{step:sa_synthesis_V_int}\quad \quad \quad Set $V_{in}= \hat T_{\mathsf x}V_v(: ~\!,k+1)$							\COMMENT{$V_{in}$ has the dimension of $(n_x\times n_\nu\times n_w,1)$} 
			\STATE \quad \quad {\bf else} 
			\STATE \quad \quad \quad Set $V_{in}= \hat T_{0\mathsf x} + \hat T_{\mathsf x}V_v(:~ \!,k+1)$ 	\COMMENT{$V_{in}$ has the dimension of $(n_x\times n_\nu\times n_w,1)$} 
			\STATE \quad \quad {\bf end}
			\STATE \label{step:sa_synthesis_min_wrt_W}\quad \quad  Reshape $V_{in}$ to a matrix $\bar V_{in}$ of the dimension $(n_x\times n_\nu,n_w)$
			\STATE \label{step_1:sa_synthesis_min_wrt_W}\quad \quad  Minimize $\bar V_{in}$ with respect to disturbance set $\hat W$ as $V_{min}$ 
			\STATE \quad \quad  Reshape $V_{min}$ to a matrix $\bar V_{min}$ of the dimension $(n_x,n_\nu)$
			\STATE \quad  \quad Maximize $\bar V_{min}$ with respect to input set $\hat U$ as $V_{max}$ of the dimension $(n_x,1)$
			\STATE \quad \quad Update $V_v(:,k) := V_{max}$
			\STATE {\bf end}	
		\end{algorithmic}
	\end{center}
\end{algorithm}

The serial algorithm does, repetitively, matrix multiplications in each loop that corresponds to different time instance of the bounded time $T_d$.
We cannot parallelize the for-loop in Step \ref{step:sa_synthesis_for_time} over time-steps due to the data dependency, however, we can parallelize the contents of this loop by simply considering standard parallel algorithms for the matrix multiplication.

\begin{algorithm}[ht]
	\caption{\small Proposed \emph{parallel} algorithm for controller synthesis satisfying safety, reachability and reach-avoid specifications} 
	\label{alg:parallel_synthesis}
	\begin{center}
		\begin{algorithmic}[1]
			\REQUIRE $\hat X$, $\hat U$, $\hat W$, bounded time horizon $T_d$, $specs \in \{ Safety,Reachability,$ $ Reach$-$Avoid \}$, target set $\mathcal{T}$ (in case $specs = Reachability, Reach$-$Avoid$), and avoid set $\mathcal{A}$ (in case $specs = Reach$-$Avoid$).
			\ENSURE Optimal satisfaction probability $V_v$ at time step $T_d = 1$, and optimal policy $\nu^\star$ corresponding to the optimal satisfaction probability.
			
			\STATE \label{step:pa_synthesis_mem_1} Compute $\hat T_{\mathsf x}$ in parallel as presented in Algorithm \ref{alg:parallell_T_computation}.
			
			\STATE {\bf if} $specs == Safety$ {\bf do} 
			\STATE \quad \quad Set value function $V_v := ones(n_x,T_d+1)$
			\STATE {\bf else}
			\STATE \label{step:pa_synthesis_mem_2} \quad \quad Compute a transition probability matrix $\hat T_{0\mathsf x}$ from $\hat X \backslash (\mathcal{T} \cup \mathcal{A})$ to $\mathcal{T}$
			\STATE \quad \quad Set $\hat T_{\mathsf x}$ to zero for any post-state in $(\mathcal{T} \cup \mathcal{A})$.
			\STATE \quad \quad Set value function $V_v := zeros(n_x,T_d+1)$
			\STATE {\bf end}
			
			\STATE {\bf for}  $k = T_d:-1:1$ (\emph{backward in time}) {\bf do} 	
			\STATE \label{step:ps_for_xu_synthesize} \quad {\bf for all} $(\bar x, \bar \nu) \in \hat X \times \hat U$ {\bf \emph{in parallel} do}
			\STATE \quad \label{step:ps_for_w_synthesize} \quad {\bf for all} $\bar w \in \hat W$
			\STATE \quad \quad \quad Construct $\hat{X}^{\mathsf{\Sigma}}_{\gamma}$ as discussed in Subsection \ref{ssec:MDP_improve_cutting_threshold}
			\STATE \label{step:pa_synthesis_comp_1} \quad \quad \quad Set $V_{in}(\bar x, \bar \nu, \bar w) := \underset{x^* \in \hat{X}^{\mathsf{\Sigma}}_{\gamma}}{\sum} V_v(x^*, k+1)T_{\mathsf x} (x^*|\bar x,\bar \nu, \bar w)$
			\STATE \quad \quad \quad {\bf if} $specs == Reach$-$Avoid$ and $\bar x \not \in (\mathcal{T} \cup \mathcal{A})$ {\bf do} 
			\STATE \label{step:pa_synthesis_comp_2} \quad \quad \quad \quad Set $V_{in}(\bar x, \bar \nu, \bar w) := V_{in}(\bar x, \bar \nu, \bar w) + T_{0\mathsf x}(\bar x, \bar \nu, \bar w)$
			\STATE \quad \quad \quad {\bf end}
			\STATE \quad \quad {\bf end} 					
			\STATE \quad {\bf end} 
			\STATE \label{step:ps_for_x_collect} \quad {\bf for all} $\bar x \in \hat X$ {\bf \emph{in parallel} do}
			\STATE \quad \quad Set $V_v(\bar x, k) := \underset{\bar \nu \in \hat U}{\max}\{ \underset{\bar w \in \hat W}{\min} \{ V_{in}(\bar x, \bar \nu, \bar w) \} \}$
			\STATE \quad \quad Set $\nu^\star(\bar x, k) := \underset{\bar \nu \in \hat U}{ \text{argmax}}\{ \underset{\bar w \in \hat W}{\min} \{  V_{in}(\bar x, \bar \nu, \bar w) \} \}$
			\STATE \quad {\bf end} 
			\STATE {\bf end}	
		\end{algorithmic}
	\end{center}
\end{algorithm}

Algorithm \ref{alg:parallel_synthesis} is a parallelization of Algorithm \ref{alg:serial_synthesis}.
Step \ref{step:ps_for_xu_synthesize} in Algorithm \ref{alg:parallel_synthesis} is the parallel implementation of the matrix multiplication in Algorithm \ref{alg:serial_synthesis}.
Step \ref{step:ps_for_x_collect} in Algorithm \ref{alg:parallel_synthesis} selects and stores the input $\bar{\nu}$ that maximizes the probabilities of enforcing the specifications. 

A significant reduction in the computation of the intermediate matrix $V_{in}$ is also introduced in Algorithm \ref{alg:parallel_synthesis}.
In Algorithm \ref{alg:serial_synthesis}, Step \ref{step:sa_synthesis_V_int}, the computation of $V_{in}$ requires a matrix multiplication between $T_{\mathsf x}$ (with a dimension of $(n_x\times n_\nu\times n_w, n_x)$) and $V_v(:,\cdot)$ (with a dimension of $(n_x, 1)$).
On the other hand, in the parallel version in Algorithm \ref{alg:parallel_synthesis}, for each $\bar w$, the corresponding computation is done for $V_{in}$ such that each element, \emph{i.e.,} $V_{in}(\bar x, \bar \nu, \bar w)$, requires only $\vert \hat{X}^{\mathsf{\Sigma}}_{\gamma} \vert$ scalar multiplications.
Here, we clearly utilize the technique discussed in Subsection \ref{ssec:MDP_improve_cutting_threshold} to consider only those post states in the cutting region $\hat{X}^{\mathsf{\Sigma}}_{\gamma}$.
Remember that other post states outside $\hat{X}^{\mathsf{\Sigma}}_{\gamma}$ are considered to have the probability zero which means we can avoid their scalar multiplications.

\subsection{On-the-Fly Construction of $\hat T_{\mathsf x}$}
In AMYTISS, we also use another technique that further reduces the required memory for computing $\hat T_{\mathsf x}$.
We refer to this approach as \emph{on-the-fly abstractions} (OFA).
In OFA version of Algorithm \ref{alg:parallel_synthesis}, we skip computing and storing the MDP $\hat T_{\mathsf x}$ and the matrix $\hat T_{0\mathsf x}$ (\emph{i.e.,} Steps \ref{step:pa_synthesis_mem_1} and \ref{step:pa_synthesis_mem_2}).
We instead compute the required entries of $\hat T_{\mathsf x}$ and $\hat T_{0\mathsf x}$ on-the-fly as they are needed (\emph{i.e.,} Steps \ref{step:pa_synthesis_comp_1} and \ref{step:pa_synthesis_comp_2}).
This significantly reduces the required memory for $\hat T_{\mathsf x}$ and $\hat T_{0\mathsf x}$ but at the cost of
repeated computation of their entries in each time step from $1$ to $T_d$. This gives the user an additional control over the trade-off between the computation time and memory.

\subsection{Supporting Multiplicative Noises and Practical Distributions}
\textsf{AMYTISS} natively supports multiplicative noises and practical distributions such as uniform, exponential, and beta distributions. The technique introduced in Subsection \ref{ssec:MDP_improve_cutting_threshold} for reducing the memory usage is also tuned for other distributions based on the support of their PDFs.
Since \textsf{AMYTISS} is designed for extensibility, it allows also for customized distributions.
Users need to specify their desired PDFs and hyper-rectangles enclosing their supports so that \textsf{AMYTISS} can include them in the parallel computation of $\hat T_{\mathsf x}$.
Further details on specifying customized distributions are provided in the README file.

\textsf{AMYTISS} also supports multiplicative noises as introduced in \eqref{Eq_1a}.
Currently, the memory reduction technique of Subsection \ref{ssec:MDP_improve_cutting_threshold} is disabled for systems with multiplicative noises.
This means users should expect larger memory requirements for systems with multiplicative noises.
However, users can still benefit from the proposed OFA version to compensate for the increase in memory requirement.
We plan to include this feature for multiplicative noises in a future update of \textsf{AMYTISS}. Note that for a better demonstration, previous sections were presented by the additive noise and Gaussian normal PDF to introduce the concepts.

\section{\textsf{AMYTISS} by Example}\label{Illustrative_example}	
\textsf{AMYTISS} is self-contained and requires only a modern \textsf{C++} compiler.
It supports the three major operating systems: \textsf{Windows}, \textsf{Linux} and \textsf{Mac OS}.
We tested \textsf{AMYTISS} on \textsf{Windows 10 x64}, \textsf{MacOS Mojave}, \textsf{Ubuntu 16.04}, and \textsf{Ubuntu 18.04}, and found no major computation time differences. Once compiled, utilizing \textsf{AMYTISS} is a matter of providing text configuration files and launching the tool.
Please refer to the provided \textsf{README} file in the repository of \textsf{AMYTISS} for a general installation instruction. 

For the sake of better illustrating the proposed algorithms and the usage of \textsf{AMYTISS}, we first introduce a simple 2-dimensional example.
Consider a robot described by the following difference equations:
\begin{equation}
\label{eq:robot_sys}
\begin{bmatrix} 
x_1(k+1)\\ 
x_2(k+1)\\  				
\end{bmatrix}
=	
\begin{bmatrix} 
x_1(k) + \tau \nu_1(k)\text{cos}(\nu_2(k)) + w(k) + \varsigma_1(k)\\ 
x_2(k) + \tau \nu_2(k)\text{sin}(\nu_2(k))+ w(k) + \varsigma_2(k)\\
\end{bmatrix}\!\!,
\end{equation}
where $(x_1, x_2) \in X := [-10,10]^2$ is a state vector representing a spacial coordinate, 
$(\nu_1, \nu_2) \in U := [-1,1]^2$ is an input vector,
$w \in W := [-1,1]$ is a disturbance,
$(\varsigma_1, \varsigma_2)$ is a noise following a Gaussian distribution with the covariance matrix 
$\mathsf{\Sigma} := \textsf{diag}(0.75, 0.75)$\footnote[3]{$\textsf{diag}(d)$ builds an $n \times n$ diagonal matrix from a supplied $n$-dimensional vector $d$.}, and
$\tau := 10$ is a constant.

To construct MDPs approximating the system, we consider a state quantization parameter of $(0.5,0.5)$, an input quantization parameter of $(0.1,0.1)$, a disturbance quantization parameter of $0.2$, and a cutting probability threshold $\gamma$ of $0.001$.
Using such quantization parameters, the number of state-input pairs $\vert \hat X \times \hat U \vert$ in $\widehat \Sigma$ is 203401.
We use $\vert \hat X \times \hat U \vert$ as an indicator of the size of the system.

System descriptions and controller synthesis requirements are provided to \textsf{AMYTISS} as text configuration files. 
The configuration files of this example is located in the directory \textsf{\%AMYTISS\%/examples/ex\_toy\_XXXX}, 
where \textsf{\%AMYTISS\%} is the installation directory of \textsf{AMYTISS} and \textsf{XXXX} should be replaced by
the controller synthesis specification of interest and can be any of: \textsf{safety}, \textsf{reachability}, or \textsf{reach-avoid}.
For a detailed description of the key-value pairs in each configuration file, refer to the \textsf{README} file in the repository of \textsf{AMYTISS}.

\subsection{Synthesis for Safety Specifications}
We synthesize a controller for the robot system in \eqref{eq:robot_sys} to keep the state of the robot inside $X$ within 8 time steps.
The synthesized controller should enforce the safety specification in the presence of the disturbance and noise.
The corresponding configuration file is located in file \textsf{\%AMYTISS\%/examples/ex\_toy\_safety/toy2d.cfg}, which describes the system
in \eqref{eq:robot_sys} and its safety requirement.
To launch \textsf{AMYTISS} and run it for synthesizing the safety controller for this example, navigate to the install directory \textsf{\%AMYTISS\%}
and run the command:
\begin{lstlisting}[frame=none]
$ pfaces -CGH -d 1 -k amytiss.cpu@./kernel-pack -cfg ./examples/ex_toy_safety/toy2d.cfg -p
\end{lstlisting}
where \lstinline{pfaces} calls \textsf{pFaces}, 
\lstinline{-CGH -d 1} asks \textsf{pFaces} to consider the first device from all CPU, GPU and HWA devices,
\lstinline{-k amytiss.cpu@./kernel-pack} asks \textsf{pFaces} to launch \textsf{AMYTISS}'s kernel from its main source folder,
\lstinline{-cfg ./examples/ex_toy} \lstinline{_safety/toy2d.cfg} asks \textsf{pFaces} to hand the configuration file to \textsf{AMYTISS}, and
\lstinline{-p} asks \textsf{pFaces} to collect profiling information.			

\begin{figure}[ht]
	\centering
	\includegraphics[width=0.55\textwidth]{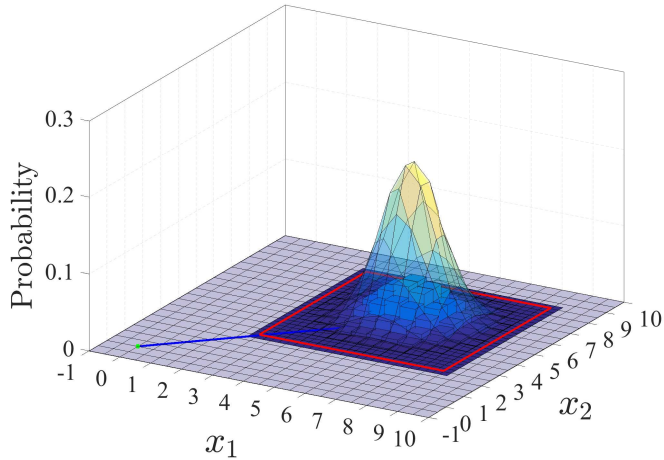}		
	\caption{
		\small A visualization of transitions for one source state $x := (0,0)$ and input $\nu = (0.7,0.8)$ of the MDP approximating the robot example. The green point is the source state, the transparent bell-like shape is the PDF and the red rectangle is the cutting region. Probabilities of reaching the partition elements inside the cutting regions are shown as bars below the PDF.
	}
	\label{fig:ex_toy_mdp}			
\end{figure}

This launches \textsf{AMYTISS} to construct an MDP of the robot system and synthesize a safety controller for it.
The results are stored in an output file specified in the configuration file.
Using the provided \textsf{MATLAB} interface in \textsf{AMYTISS}, we visualize some transitions of the constructed MDP and show them in Figure \ref{fig:ex_toy_mdp}.
The used \textsf{MATLAB} script is located in \textsf{\%AMYTISS\%/examples/ex\_toy\_safety/make\_figs.m}.

\begin{figure}[ht]
	\hspace*{-1.5cm}
	\centering
	\includegraphics[width=0.9\textwidth]{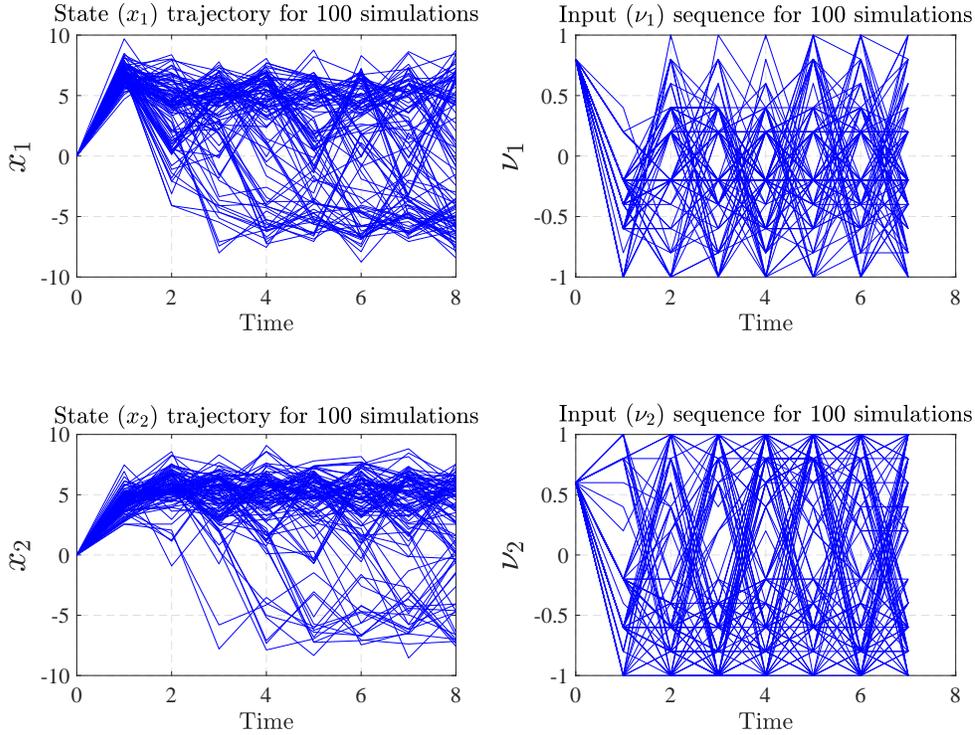}
	\caption{
		\small 100 different simulations of the closed-loop behavior of the robot under a safety controller synthesized for maintaining the robot inside $X$.
		At left, we show the state trajectory of the system at each time step.
		At right, we show the applied input at each time step. 
		For the sake of readability, the input plots are shown as piece-wise linear signal.}\vspace{0.1cm}
	\label{fig:ex_toy_sim}			
\end{figure}

The output file contains also the control strategy which we use to simulate the closed-loop behavior of the system.
Again, we rely on the the provided \textsf{MATLAB} interface in \textsf{AMYTISS} to simulate the closed-loop behavior.
The \textsf{MATLAB} script in \textsf{\%AMYTISS\%/examples/ex\_toy\_safety/closedloop.m} simulates the system with random choices on $\bar w \in \hat W$ and random values for the noise according to the given covariance matrix.
At each time step, the simulation queries the strategy from the output file and applies it to the system.
We repeat the simulation 100 times.
Figure \ref{fig:ex_toy_sim} shows the closed-loop simulation results.
Note that the input is always fixed at the time step $k = 0$.
This is because we store only one input, which is the one maximizing the probability of satisfaction.
After the time step $k = 0$ and due to the noise/disturbance, the system lands in different states which requires applying different inputs to satisfy the specification.

	\subsection{Synthesis for Reach-Avoid Specifications}			

\begin{figure}[ht]
	\centering
	\includegraphics[width=0.6\textwidth]{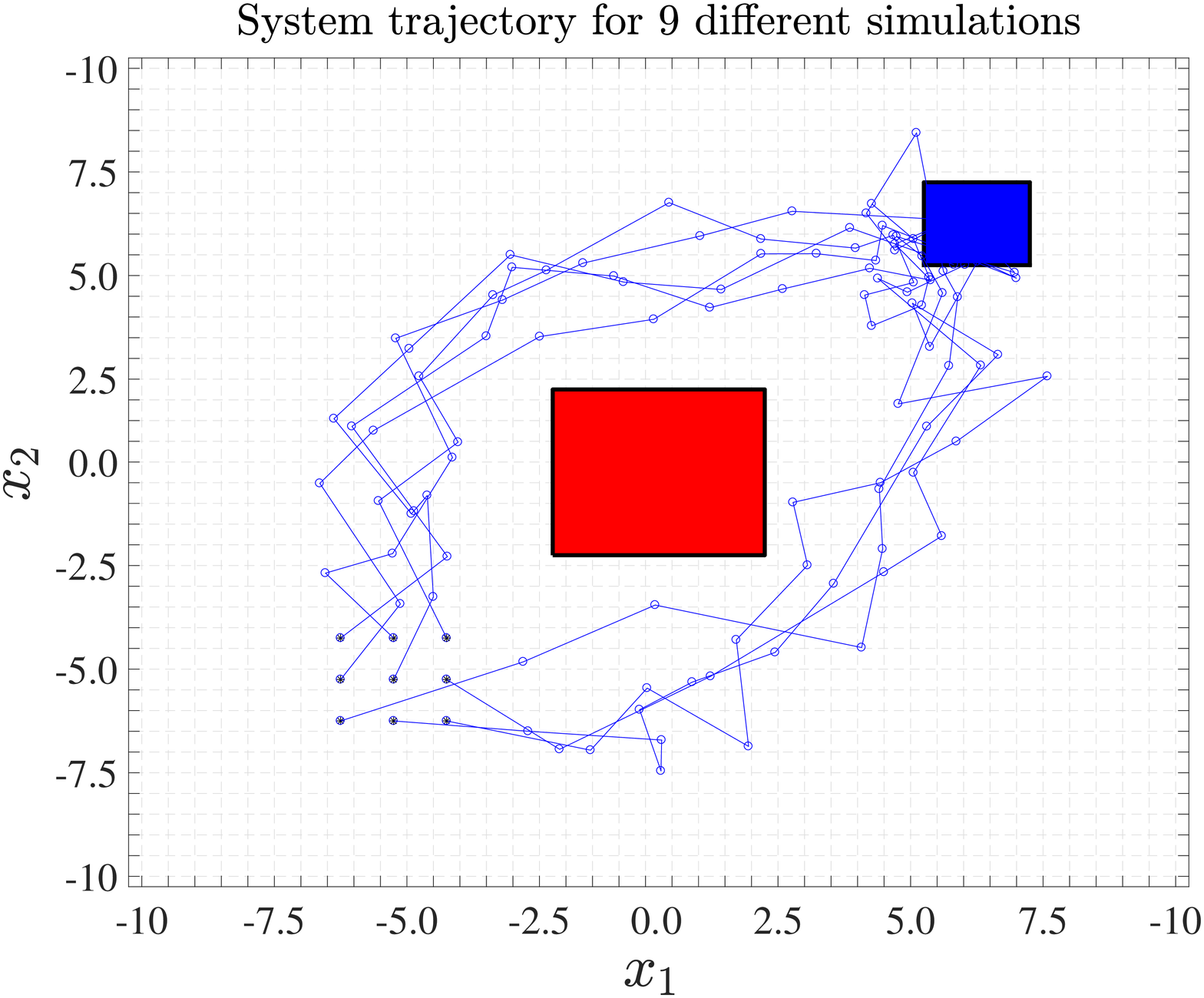}
	\caption{
		\small 9 different simulations of the closed-loop behavior of the robot example under a synthesized controller enforcing the robot to reach a target set while avoiding an avoid set.
		The 9 dots at the left bottom correspond to 9 initial states for 9 different simulation runs.
		The red rectangle is the avoid set of states.
		The blue rectangle is the target set of states.
	}
	\label{fig:ex_toy_sim_reachavoid}			
\end{figure}

We synthesize a controller for the robot system in \eqref{eq:robot_sys} to reach the set $[5,7]^2$ while avoiding the set $[-2,2]^2$ within 16 time steps.
To launch \textsf{AMYTISS} and run it for synthesizing the reach-avoid controller for this example, navigate to the install directory \textsf{\%AMYTISS\%}
and run the command:
\begin{lstlisting}[frame=none]
$ pfaces -CGH -k amytiss.cpu@./kernel-pack -cfg ./examples/ex_toy_reachavoid/toy2d.cfg -d 1 -p
\end{lstlisting}
This launches \textsf{AMYTISS} to construct an MDP of the robot system and synthesize a reach-avoid controller for it. A \textsf{MATLAB} script simulates the closed-loop and it is located in \textsf{\%AMYTISS\%/examples/ex\_toy\_reachavoid/closed}\\\textsf{loop.m}.
This runs 9 different simulations from 9 different initial states.
Figure \ref{fig:ex_toy_sim_reachavoid} shows the closed-loop simulation results.

\section{Benchmarking and Case Studies}\label{Case_Study}

\subsection{Controlling the Computational Complexities}
\label{controlling_complexities}		

\textsf{AMYTISS} implements scalable parallel algorithms that run on top of \textsf{pFaces}.
Hence, users can utilize computing power in HPC platforms and cloud computing to scale the computation and control the computational complexities of their problems.
We fix the system (\emph{i.e.,} the robot example) in hand and show how \textsf{AMYTISS} scales with respect to different computing platforms.
Table \ref{tbl:HW-configurations} lists the HW configuration we use to benchmark \textsf{AMYTISS}.
The devices range from local devices in laptops and desktop computers to advanced compute devices in Amazon AWS cloud computing services.
\vspace{0.3cm}
\begin{table}[ht]
	\small
	\centering
	\caption{\small HW configurations for benchmarking \textsf{AMYTISS}.}
	\vspace{2mm}
	\begin{tabular}{ p{1.2cm} | p{8.1cm} | p{0.7cm} | p{1.7cm}}
		\textbf{Id} & 
		\textbf{Description} &
		\textbf{PEs} &
		\textbf{Frequency}
		\\ \hline
		
		\textbf{CPU$_1$}
		& Local machine: Intel Xeon E5-1620
		& 8
		& 3.6 GHz
		\\ \hline																
		
		\textbf{CPU$_2$}
		& Macbook Pro 15: Intel i9-8950HK
		& 12
		& 2.9 GHz
		\\ \hline

		\textbf{CPU$_3$}
		& \specialcell{AWS instance {\tt c5.18xlarge}: Intel Xeon Platinum 8000}
		& 72	
		& 3.6 GHz
		\\ \hline
		
		\textbf{GPU$_1$}
		& \specialcell{Macbook Pro 15 laptop laptop: Intel UHD Graphics 630}
		& 23	
		& 0.35 GHz
		\\ \hline							
		
		\textbf{GPU$_2$}
		& \specialcell{Macbook Pro 15 laptop: AMD Radeon Pro Vega 20}
		& 1280	
		& 1.2 GHz
		\\ \hline										
		
		\textbf{GPU$_3$}
		& \specialcell{AWS p3.2xlarge instance: NVIDIA Tesla V100}
		& 5120 	
		& 0.8 GHz
		\\ \hline										
		
	\end{tabular}\vspace{0.2cm}
	\label{tbl:HW-configurations}
\end{table}	

Table~\ref{benchmark_Comparion} shows the benchmarking results running \textsf{AMYTISS} with these 
HWCs for several case studies and makes comparisons between \textsf{AMYTISS}, 
\textsf{FAUST}$^{\mathsf 2}$, and \textsf{StocHy}. We employ a machine with Windows operating system (Intel i7@3.6GHz CPU and 16 GB of RAM) for \textsf{FAUST}$^{\mathsf 2}$, and \textsf{StocHy}. It should be mentioned that \textsf{FAUST}$^{\mathsf 2}$ predefines a minimum number of representative points 
based on the desired abstraction error, and accordingly the computation time and memory usage reported in Table~\ref{benchmark_Comparion} are based on the minimum number of representative points. In addition, to have a fair comparison, we run all the case studies with additive noises since neither \textsf{FAUST}$^{\mathsf 2}$ nor \textsf{StocHy} supports multiplicative noises.

For each HWC, we show the time in seconds to solve the problem.
Clearly, employing HWCs with more PEs reduces the time to solve the problem.
This is a strong indication for the scalability of the proposed algorithms.
This also becomes very useful in real-time applications, where users can control the computation time of their problems by adding more resources.
Since, \textsf{AMYTISS} is the only tool that can utilize the reported HWCs, we do not compare it with other similar tools.

To show the applicability of our results to large-scale stochastic systems, we apply our proposed techniques to several physical case studies.
First, we synthesize a controller for $3$- and $5$-dimensional \emph{room temperature networks} to keep temperature of rooms in a comfort zone. Then we synthesize a controller for \emph{road traffic networks} with $3$ and $5$ dimensions to keep the density of the traffic below some level. 
We then consider $3$- and $7$-dimensional \emph{nonlinear} models of an autonomous vehicle and synthesize reach-avoid controllers to automatically park the vehicles. 
For each case study, we compare our tool with \textsf{FAUST}$^{\mathsf 2}$ and \textsf{StocHy} and report the technical details in Table~\ref{benchmark_Comparion}.

\subsection{Room Temperature Network}
\subsubsection{5-Dimensional System.}
We first apply our results to the temperature regulation of $5$ rooms each equipped with a heater and connected on a circle. The model of this case study is borrowed from \cite{lavaei2017HSCC}. The evolution of temperatures $T_{x_i}$ 
can be described by individual rooms as
\begin{equation*}
\Sigma_{\mathfrak a_i}:\left\{\hspace{-1mm}\begin{array}{l}T_{x_i}(k+1)=a_{ii}T_{x_i}(k)+\gamma T_{h} \nu_i(k)+\eta w_i(k)+\beta T_{ei}+0.01\varsigma_i(k), i\in\{1,3\},\\
T_{x_i}(k+1)=b_{ii}T_{x_i}(k)+\eta w_i(k)+\beta T_{ei}+0.01\varsigma_i(k), i\in\{2,4,5\},\\
y_i(k)=T_{x_i}(k),\\
\end{array}\right.
\end{equation*}
where $a_{ii}=(1-2\eta-\beta-\gamma\nu_{i}(k))$, $b_{ii}=(1-2\eta-\beta)$, and $w_i(k) = T_{x_{i-1}}(k) + T_{x_{i+1}}(k)$ (with $T_{x_{0}} = T_{x_{n}}$ and $T_{x_{n+1}} = T_{x_{1}}$).
Parameters $\eta = 0.3$, $\beta = 0.022$, and $\gamma = 0.05$ are conduction factors, respectively, between rooms $i \pm 1$ and the room $i$, between the external environment and the room $i$, and between the heater and the room $i$. Moreover, $T_{ei}=-1\,^\circ C$, $T_h=50\,^\circ C$ are outside and heater temperatures, and $T_i(k)$ and $\nu_i(k)$ are taking values in sets $[19,21]$ and $[0,1]$, respectively, $\forall i\in\{1,\ldots,n\}$.

Let us now synthesize a controller for $\Sigma_\mathfrak a$ via the abstraction $\widehat \Sigma_\mathfrak a$ such that the controller maintains the temperature of any room in the safe set $[19,21]$ for at least $8$ time steps.

\subsubsection{3-Dimensional System.} We also apply our algorithms to a smaller version of this case study ($3$-dimensional system) with the results reported in Table~\ref{benchmark_Comparion}.
\vspace{0.5cm}

\subsection{Road Traffic Network}

\subsubsection{5-Dimensional System.}\label{ex_traffic}	

\begin{figure}[ht]
	\begin{center}
		\includegraphics[width=5cm]{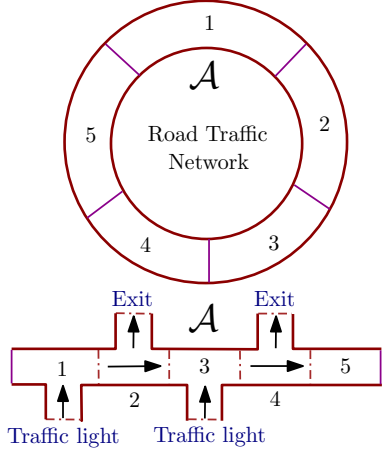}
		\caption{\small Model of a road traffic network composed of $5$ cells of $500$ meters with $2$ entries and $2$ ways out.}
		\label{Fig2}
	\end{center}
\end{figure}

Consider a road traffic network divided in $5$ cells of $500$ meters with $2$ entries and $2$ ways out, 
as schematically depicted in Figure \ref{Fig2}. 
The model of this case study is borrowed from \cite{le2013mode} by including the stochasticity in the model as the additive noise.

The two entries are controlled by traffic lights, denoted by $\nu_1$ and $\nu_3$, that enable (green
light) or not (red light) the vehicles to pass. In this model, the length of a cell is in kilometers
[km] and the flow speed of vehicles is $100$ kilometers per hour [km/h]. Moreover,
during the sampling time interval $\tau = 6.48$ seconds, it is assumed that $6$ vehicles pass the entry controlled by
the light $\nu_1$, $8$ vehicles pass the entry controlled by the light $\nu_3$, and one quarter of vehicles that leave cells 
$1$ and $3$ goes out on the first exit (its ratio denoted by $q$). We want to observe the density of the traffic $x_i$, given in 
vehicles per cell, for each cell $i$ of the road. The model of cells is described by:
\begin{align}\notag
x_1(k+1) &= (1-\frac{\tau v_1}{L_1})x_1(k) + \frac{\tau v_{5}}{L_{5}} w_1(k)+6\nu_1(k)+0.7\varsigma_1(k), \\\notag
x_i(k+1) &= (1-\frac{\tau v_i}{L_i}-q)x_i(k) + \frac{\tau v_{i-1}}{L_{i-1}} w_i(k)+0.7\varsigma_i(k), \quad i \in \{2,4\},\\\notag
x_3(k+1) &= (1-\frac{\tau v_3}{L_3})x_3(k) + \frac{\tau v_{2}}{L_{2}} w_3(k)+8\nu_3(k)+0.7\varsigma_3(k), \\\notag
x_5(k+1) &= (1-\frac{\tau v_5}{L_5})x_5(k) + \frac{\tau v_{4}}{L_{4}} w_5(k)+0.7\varsigma_5(k),
\end{align}
where $w_i(k) = x_{i-1}(k)$ (with $x_0 = x_{5}$).
We are interested first in constructing the finite MDP of the given $5$-dimensional system and then synthesizing policies keeping the density of the traffic lower than $10$ vehicles per cell.

For this example, 
we have $X := [0,10]^5$ with a quantization parameter of $(0.37,0.37,0.37,0.37,0.37)$,
$U = [0,1]^2$ with a quantization parameter of $(1,1)$,
a noise covariance matrix $\mathsf{\Sigma} := \textsf{diag}(0.7, 0.7, 0.7, 0.7, 0.7)$,
and a cutting probability level $\gamma$ of $2e-2$.

\subsubsection{3-Dimensional System.} We also apply our algorithms to the same case study but with  $3$ dimensions  for the sake of benchmarking.

\subsection{Autonomous Vehicle}\label{ex_vehicle}

\subsubsection{7-Dimensional BMW $320$i.} Here, to show the applicability of our approaches to \emph{nonlinear} models, 
we consider a vehicle described by the following hybrid $7$-dimensional \emph{nonlinear} single track (ST) model of a BMW $320$i car \cite[Section 5.1]{Althof18} by including the stochasticity inside the dynamics as the additive noise:

For $\vert x_4(k) \vert < 0.1$:
\begin{align}\notag
x_i(k+1) &= x_i(k) + \tau a_i(k) + R_i\varsigma_i(k), \quad i \in \{1,\dots, 7\}\backslash\{3,4\}, \\\notag
x_3(k+1) &= x_3(k) + \tau \text{Sat}_1(\nu_1) + 0.2\varsigma_3(k),\\\notag
x_4(k+1) &= x_4(k) + \tau \text{Sat}_2(\nu_2) + 0.1\varsigma_4(k),
\end{align}
for $\vert x_4(k) \vert \ge 0.1$:
\begin{align}\notag
x_i(k+1) &= x_i(k) + \tau b_i(k) + R_i\varsigma_i(k), \quad i \in \{1,\dots, 7\}\backslash\{3,4\}, \\\notag
x_3(k+1) &= x_3(k) + \tau \text{Sat}_1(\nu_1) + 0.2\varsigma_3(k),\\\notag
x_4(k+1) &= x_4(k) + \tau \text{Sat}_2(\nu_2) + 0.1\varsigma_4(k),
\end{align}			

where,
\begin{align}\notag
R_1 &= R_2 = 0.25, \quad R_5 = R_6 = R_7 = 0.2,\quad a_1 = x_4\text{cos}(x_5(k)),\quad a_2 = x_4\text{sin}(x_5(k)),\\\notag
\quad a_5 &= \frac{x_4}{l_{wb}} \text{tan}(x_3(k)),\quad a_6 = \frac{\nu_2(k)}{l_{wb}} \text{tan}(x_3(k)) + \frac{x_4}{l_{wb}\text{cos}^2(x_3(k))} \nu_1(k), \quad a_7 = 0,\\\notag
b_1 &= x_4(k)\text{cos}(x_5(k)+x_7(k)),\quad b_2 =x_4(k)\text{sin}(x_5(k)+x_7(k)),\quad b_5 = x_6(k),\\\notag
b_6 &= \frac{\bar\mu m}{I_z(l_r+l_f)}(l_fC_{S,f}(gl_r-\nu_2(k)h_{cg})x_3(k)+(l_rC_{S,r}(gl_f+\nu_2(k)h_{cg})-l_fC_{S,f}(gl_r-\nu_2(k)h_{cg}))x_7(k)\\\notag
&\quad-(l_f^2C_{S,f}(gl_r-\nu_2(k)h_{cg})+l_r^2C_{S,r}(gl_f+\nu_2(k)h_{cg}))\frac{x_6(k)}{x_4(k)}),\\\notag
b_7 &= \frac{\bar\mu_f}{x_4(k)(l_r+l_f)}(C_{S,f}(gl_r-\nu_2(k)h_{cg})x_3(k)+(C_{S,r}(gl_f+\nu_2(k)h_{cg})+C_{S,f}(gl_r-\nu_2(k)h_{cg}))x_7(k)\\\notag
&\quad-(l_fC_{S,f}(gl_r-\nu_2(k)h_{cg})-l_rC_{S,r}(gl_f+\nu_2(k)h_{cg}))\frac{x_6(k)}{x_4(k)})-x_6(k).
\end{align}
Here, $\text{Sat}_1(\cdot)$ and $\text{Sat}_2(\cdot)$ are input saturation functions introduced in \cite[Section 5.1]{Althof18},
$x_1$ and $x_2$ are the position coordinates, 
$x_3$ is the steering angle, 
$x_4$ is the heading velocity, 
$x_5$ is the yaw angle, 
$x_6$ is the yaw rate, and 
$x_7$ is the slip angle. 
Variables $\nu_1$ and $\nu_2$ are inputs and they control the steering angle and heading velocity, respectively. 

The model takes into account the tire slip making it a good candidate for studies that consider planning of evasive maneuvers that are very close to physical limits.  We consider an update period $\tau = 0.1$ seconds and the following parameters for a BMW $320$i car: 
$l_{wb} = 2.5789$ as the wheelbase,
$m = 1093.3$ [kg] as the total mass of the vehicle,
$\bar\mu = 1.0489$ as the friction coefficient, 
$l_f = 1.156$ [m] as the distance from the front axle to the center of gravity (CoG), 
$l_r = 1.422$ [m] as the distance from the rear axle to CoG, 
$h_{cg} = 0.6137$ [m] as the hight of CoG, 
$I_z = 1791.6$ [kg m$^2$] as the moment of inertia for entire mass around $z$ axis, 
$C_{S,f} = 20.89$ [$1$/rad] as the front cornering stiffness coefficient, and 
$C_{S,r} = 20.89$ [$1$/rad] as the rear
cornering stiffness coefficient.

To construct a finite MDP $\widehat\Sigma_\mathfrak a$, we consider a bounded version of the state
set $X := [-10.0,10.0] \times [-10.0,10.0]\times [-0.40,0.40]\times[-2,2]\times[-0.3,0.3]\times[-0.4,0.4]\times[-0.04, 0.04]$, 
a state discretization vector $[4.0;4.0;0.2;1.0;0.1;0.2;\\0.02]$, an input
set $U := [-0.4,0.4]\times[-4,4]$, and an input discretization vector $[0.2;2.0]$.

We are interested in an autonomous operation of the vehicle. 
The vehicle should park itself automatically in the parking lot located in the projected set $[-1.5,0.0] \times [0.0,1.5]$ within 32 time steps.
The vehicle should avoid hitting a barrier represented  by the set $[-1.5,0.0] \times [-0.5,0.0]$.

\subsubsection{3-Dimensional Autonomous Vehicle.} We also apply our algorithms to a $3$-dimensional autonomous vehicle \cite[Section IX-A]{reissig2016feedback} for the sake of benchmarking.

\subsection{Benchmark in \textsf{StocHy}}
\label{ssec_compare_Wth_stochy}
We benchmark our results against the ones provided by \textsf{StocHy} \cite{StocHy19}. 
We employ the same case study as in \cite[Case study 3]{StocHy19} which starts from 2-dimensional to $12$-dimensional continuous-space systems with the same parameters. 

To have a fair comparison, we utilize a machine with the same configuration as the one employed in \cite{StocHy19} (a laptop having an Intel Core i$7-8550$U CPU at $1.80$GHz with $8$ GB of RAM). 
We build a finite MDP for the given model and compare our computation time with the results provided by \textsf{StocHy}.

Table~\ref{tbl:compare_with_stochy} shows the comparison between \textsf{StocHy} and \textsf{AMYTISS}.
\textsf{StocHy} suffers significantly from the state-explosion problem as seen from its exponentially growing computation time.
\textsf{AMYTISS}, on the other hand, outperforms \textsf{StocHy} and can handle bigger systems using the same hardware. This comparison shows speedups up to maximum $375$ times for the $12$-dimensional system.
Note that we only reported up to 12-dimensions but \textsf{AMYTISS} can readily go beyond this limit for this example.
For instance, \textsf{AMYTISS} managed to handle the 20-dimensional version of this system in 1572 seconds using an NVIDIA Tesla V100 GPU in Amazon AWS.

\begin{table}[ht]
	\caption{
		\small Comparison between \textsf{StocHy} and \textsf{AMYTISS} for a continuous-space system with dimensions up to 12.
		The reported system is autonomous and, hence, $\hat U$ is singleton.
		$\vert \hat X \vert$ refers to the size of the system.
	}					
	\resizebox{\columnwidth}{!}{
		\begin{tabular}{l | l | l | l | l | l | l | l | l | l | l | l }
			\textbf{Dimension}                                                                         
			& \textbf{2} & \textbf{3} & \textbf{4} & \textbf{5} & \textbf{6} & \textbf{7} & \textbf{8} & \textbf{9} & \textbf{10} & \textbf{11} & \textbf{12} \\ \hline
			
			\textbf{$\vert \hat X \vert$}                                                                           
			&  4          &  8             &    16       &  32         &  64         &       128  &   265     &   512         &   1024        & 2048     &   4096 \\  \hline
			
			\textbf{\begin{tabular}[c]{@{}l@{}}Time (s) - StocHy\end{tabular}}     
			& 0.015      & 0.08       & 0.17       & 0.54       & 2.17      & 9.57     & 40.5      & 171.6     & 385.5    & 1708.2     & 11216    \\ \hline
			
			\textbf{\begin{tabular}[c]{@{}l@{}}Time (s) - AMYTISS\end{tabular}}     
			&  0.02      & 0.92        & 0.20      & 0.47      &  1.02       & 1.95        &  3.52       &  6.32        & 10.72       &  17.12        &   29.95          \\ \hline
		\end{tabular}
	}\vspace{0.1cm}
	\label{tbl:compare_with_stochy}
\end{table}

Readers are highly advised to pay attention to the size of the system $\vert \hat X \times \hat U \vert$ (or $\vert \hat X \vert$ when $\hat U$ is singleton), not to its dimension. Actually, here, the 12-dimensional system, which has a size of 4096 state-input pairs is much smaller than the 2-dimensional illustrative 
example we introduced in Section \ref{Illustrative_example}, which has a size of 203401 state-input pairs.
The current example has a small size due to the very coarse quantization parameters and the tight bounds used to quantize $X$.

As seen in Table~\ref{benchmark_Comparion}, \textsf{AMYTISS} outperforms \textsf{FAUST}$^{\mathsf 2}$ and \textsf{StocHy}
in all the case studies (maximum speedups up to $692 000$ times).
Moreover, \textsf{AMYTISS} is the only tool that can utilize the available HW resources.
The OFA feature in \textsf{AMYTISS} reduces dramatically the required memory, while still 
solves the problems in a reasonable time.
\textsf{FAUST}$^{\mathsf 2}$ and \textsf{StocHy} fail to solve many of the problems since they lack the native support for nonlinear systems, 
they require large amounts of memory, or they do not finish computing within 24 hours.

	\begin{landscape}	
	\newcommand{\LTO}{$\le$ 1.0}
	\newcommand{\MTD}{$\ge$ 24h}
	\newcommand{\NEM}{N/M}
	\newcommand{\NA}{N/A}
	\newcommand{\NSN}{N/S}
	\newcommand{\MCNEM}{\multicolumn{2}{c|}{\NEM}}
	\newcommand{\MCNA}{\multicolumn{2}{c|}{\NA}}
	\newcommand{\MCNSN}{\multicolumn{2}{c|}{\NSN}}
	\newcommand{\MCMTD}{\multicolumn{2}{c|}{\MTD}}
	\begin{table}
		\vspace{-0.5cm}
		\small
		\centering				
		\caption{
			\small Comparison between \textsf{AMYTISS}, \textsf{FAUST}$^{\mathsf 2}$ and \textsf{StocHy} based on their native features
			for several (physical) case studies. 
			CSB refers to the continuous-space benchmark provided in \cite{StocHy19}.
			$\dagger$ refers to cases when we run \textsf{AMYTISS} with the OFA algorithm.				
			\NEM{} refers to the situation when there is not enough memory to run the case study.
			\NSN{} refers to the lack of native support for nonlinear systems.
			(Kx) refers to an $1000$-times speedup. 
			The presented speedup is the maximum speedup value across all reported devices. 
			The required memory usage and computation time for \textsf{FAUST}$^{\mathsf 2}$ and \textsf{StocHy} are reported for just constructing finite MDPs. 
			The reported times and memories are respectively in seconds and  MB, unless other units are denoted.
		}
		\vspace{2mm}
		{\footnotesize \begin{tabular}{ 
				p{2.3cm}  | l | l | l | 
				l | l | l | l | l | l | l |
				p{0.69cm} | p{0.6cm} | 
				p{0.69cm} | p{0.77cm} |
				p{1.17cm} | p{1.28cm} 
			}
			
			\toprule
			& & & &  
			\multicolumn{7}{|c|}{\textbf{\textsf{AMYTISS} (time)}} & 
			\multicolumn{2}{|c|}{\textbf{\textsf{FAUST}$^{\mathsf 2}$}} & 
			\multicolumn{2}{|c|}{\textbf{\textsf{StocHy}}} &
			\multicolumn{2}{|c}{\textbf{Speedup w.r.t}} \\
			
			\textbf{Problem} & 
			\textbf{Spec.} & 
			\textbf{$\vert \hat{X} \times \hat{U} \vert$} & 
			\textbf{$T_d$} & 
			
			\textbf{Mem.} &				
			CPU$_1$ & 
			CPU$_2$ & 
			CPU$_3$ & 
			GPU$_1$ & 
			GPU$_2$ & 
			GPU$_3$ &
			
			\textbf{Mem.} &				
			\textbf{Time} &  
			\textbf{Mem.} &				
			\textbf{Time} & 
			\textsf{FAUST} &
			\textsf{StocHy} \\
			
			\midrule
			
			\begin{tabular}[c]{@{}l@{}}$2$-d StocHy CSB \end{tabular}			&  Safety 	&  4 		& 6	& \LTO		&\LTO	&\LTO	&\LTO	&\LTO   &\LTO   &0.0001	& \LTO	& 0.002 & 8.5	& 0.015	&\textbf{20 x}& \textbf{150 x}	\\ \hline
			\begin{tabular}[c]{@{}l@{}}$3$-d StocHy CSB\end{tabular}			&  Safety 	&  8		& 6	& \LTO		&\LTO	&\LTO	&\LTO	&\LTO   &\LTO   &0.0001	& \LTO	& 0.002 & 8.5	& 0.08	&\textbf{20 x}& \textbf{800 x}\\ \hline
			\begin{tabular}[c]{@{}l@{}}$4$-d StocHy CSB \end{tabular}			&  Safety 	&  16 		& 6	& \LTO		&\LTO	&\LTO	&\LTO	&\LTO   &\LTO   &0.0002	& \LTO	& 0.01	& 8.5	& 0.17	&\textbf{50 x}& \textbf{850 Kx}\\ \hline
			\begin{tabular}[c]{@{}l@{}}$5$-d StocHy CSB\end{tabular}			&  Safety 	&  32		& 6	& \LTO		&\LTO	&\LTO	&\LTO	&\LTO   &\LTO   &0.0003	& \LTO	& 0.01	& 8.7	& 0.54	&\textbf{33 x}& \textbf{1.8 Kx}\\ \hline
			\begin{tabular}[c]{@{}l@{}}$6$-d StocHy CSB \end{tabular}			&  Safety 	&  64		& 6	& \LTO		&\LTO	&\LTO	&\LTO	&\LTO   &\LTO   &0.0006	& 4.251	& 1.2	& 9.6	& 2.17	&\textbf{2.0 Kx}& \textbf{3.6 Kx}\\ \hline
			\begin{tabular}[c]{@{}l@{}}$7$-d StocHy CSB\end{tabular}			&  Safety 	&  128		& 6	& \LTO		&\LTO	&\LTO	&\LTO	&\LTO   &\LTO   &0.0012	& 38.26	& 6	& 12.9	& 9.57	&\textbf{5 Kx}& \textbf{7.9 Kx}\\ \hline
			\begin{tabular}[c]{@{}l@{}}$8$-d StocHy CSB \end{tabular}			&  Safety 	&  256		& 6	& \LTO		&\LTO	&\LTO	&\LTO	&\LTO   &\LTO   &0.0026	& 344.3	& 37	& 26.6	& 40.5	&\textbf{14.2 Kx}& \textbf{15.6 Kx}\\ \hline
			\begin{tabular}[c]{@{}l@{}}$9$-d StocHy CSB\end{tabular}			&  Safety 	&  512		& 6	& 1.0		&\LTO	&\LTO	&\LTO	&\LTO   &\LTO   &0.0057	& 3 GB	& 501  & 80.7	& 171.6 &\textbf{87.8 Kx}& \textbf{30.1 Kx}\\ \hline
			\begin{tabular}[c]{@{}l@{}}$10$-d StocHy CSB\end{tabular}			&  Safety 	&  1024		& 6	& 4.0		&\LTO	&\LTO	&\LTO	&\LTO   &\LTO   &0.0122	& \MCNEM		& 297.5	& 385.5 & \NA	& \textbf{32 Kx}\\ \hline
			\begin{tabular}[c]{@{}l@{}}$11$-d StocHy CSB \end{tabular}			&  Safety 	&  2048		& 6	& 16.0		&1.0912	&\LTO	&\LTO	&\LTO   &\LTO   &0.0284	& \MCNEM		& 1 GB  & 1708.2& \NA	& \textbf{60 Kx}\\ \hline
			\begin{tabular}[c]{@{}l@{}}$12$-d StocHy CSB\end{tabular}			&  Safety 	&  4096		& 6	& 64.0		&4.3029	&4.1969	&\LTO	&\LTO   &\LTO   &0.0624	& \MCNEM		& 4 GB	& 11216 & \NA	& \textbf{179 Kx}\\ \hline
			\begin{tabular}[c]{@{}l@{}}$13$-d StocHy CSB\end{tabular}			&  Safety 	&  8192		& 6	& 256.0		&18.681	&19.374	&1.8515	&1.6802 &\LTO   &0.1277	& \MCNEM		& \NA   & \MTD  & \NA	& \textbf{$\ge$676 Kx}\\ \hline
			\begin{tabular}[c]{@{}l@{}}$14$-d StocHy CSB\;\;\end{tabular}		&  Safety 	&  16384	& 6	& 1024.0	&81.647	&94.750	&7.9987	&7.3489 &6.1632 &0.2739	& \MCNEM		& \NA	& \MTD	& \NA	& \textbf{$\ge$320 Kx}\\ \hline				
			\begin{tabular}[c]{@{}l@{}}$2$-d Robot$\dagger$\end{tabular}		&  Safety 	&  203401	& 8	& \LTO		&8.5299	&5.0991	&0.7572	&\LTO	&\LTO   &0.0154	& \MCNA			& \MCNA			& \NA	& \NA	\\ \hline
			\begin{tabular}[c]{@{}l@{}}$2$-d Robot\end{tabular}					&  R.Avoid 	&  741321	&16 & 482.16	&48.593	&18.554	&4.5127	&2.5311 &3.4353 &0.3083	& \MCNSN		& \MCNSN		& \NA	& \NA	\\ \hline
			\begin{tabular}[c]{@{}l@{}}$2$-d Robot$\dagger$\end{tabular}		&  R.Avoid	&  741321	&16 & 4.2484	&132.10	&41.865	&11.745	&5.3161 &3.6264 &0.1301	& \MCNA			& \MCNA			& \NA	& \NA	\\ \hline
			\begin{tabular}[c]{@{}l@{}}$3$-d Room Temp.\end{tabular}			&  Safety 	&  7776		& 8	& 6.4451	&0.1072	&0.0915	&0.0120	&\LTO   &\LTO   &0.0018	& 3.12	& 1247	& \MCNEM		&\textbf{692 Kx}& \NA	\\ \hline
			\begin{tabular}[c]{@{}l@{}}$3$-d Room Temp.$\dagger$\end{tabular}	&  Safety 	&  7776		& 8	& \LTO		&0.5701	&0.3422	&0.0627	&\LTO   &\LTO   &0.0028	& \MCNA			& \MCNA			& \NA	& \NA	\\ \hline
			\begin{tabular}[c]{@{}l@{}}$5$-d Room Temp.\end{tabular}			&  Safety 	&  279936	& 8	& 3338.4	&200.00	&107.93	&19.376	&10.084 &\NEM   &1.8663	& 2 GB	& 3248	& \MCNEM		&\textbf{1740 x}& \NA	\\ \hline
			\begin{tabular}[c]{@{}l@{}}$5$-d Room Temp.$\dagger$\end{tabular}	&  Safety 	&  279936	& 8	& 1.36		&716.84	&358.23	&63.758	&30.131 &22.334 &0.5639	& \MCNA			& \MCNA			& \NA	& \NA	\\ \hline
			\begin{tabular}[c]{@{}l@{}}$3$-d Road Traffic\end{tabular}			&  Safety 	&  2125764	&16	& 1765.7	&29.200	&131.30	&3.0508	&5.7345 &10.234 &1.2895	& \MCNEM		& \MCNEM		& \NA	& \NA	\\ \hline
			\begin{tabular}[c]{@{}l@{}}$3$-d Road Traffic$\dagger$\end{tabular}	&  Safety 	&  2125764	&16	& 14.19		&160.45	&412.79	&13.632	&12.707 &11.657 &0.3062	& \MCNA			& \MCNA			& \NA	& \NA	\\ \hline
			\begin{tabular}[c]{@{}l@{}}$5$-d Road Traffic\end{tabular}      	&  Safety 	&  68841472 & 7	& 8797.4	&\NEM	&537.91	&38.635	&\NEM   &\NEM   &4.3935	& \MCNEM		& \MCNEM		& \NA	& \NA	\\ \hline	
			\begin{tabular}[c]{@{}l@{}}$5$-d Road Traffic$\dagger$\end{tabular}	&  Safety 	&  68841472	& 7	& 393.9		&1148.5	&1525.1	&95.767	&44.285 &36.487 &0.7397	& \MCNA			& \MCNA			& \NA	& \NA	\\ \hline
			\begin{tabular}[c]{@{}l@{}}$3$-d Vehicle\end{tabular}				&  R.Avoid 	&  1528065	&32	& 1614.7	&2.5h	&1.1h	&871.89	&898.38 &271.41 &10.235	& \MCNSN		& \MCNSN		& \NA	& \NA	\\ \hline
			\begin{tabular}[c]{@{}l@{}}$3$-d Vehicle$\dagger$\end{tabular}		&  R.Avoid 	&  1528065	&32	& 11.17		&2.8h	&1.9h	&879.78	&903.2  &613.55 &107.68	& \MCNA			& \MCNA			& \NA	& \NA	\\ \hline
			\begin{tabular}[c]{@{}l@{}}$7$-d BMW $320$i\end{tabular}			&  R.Avoid	&  3937500	&32	& 10169.4	&\NEM	&\MTD	&21.5h	&\NEM   &\NEM   &825.62	& \MCNSN		& \MCNSN		& \NA	& \NA	\\ \hline
			\begin{tabular}[c]{@{}l@{}}$7$-d BMW $320$i$\dagger$\end{tabular}	&  R.Avoid	&  3937500	&32	& 30.64		&\MTD	&\MTD	&\MTD	&\MTD   &\MTD   &1251.7	& \MCNA			& \MCNA			& \NA	& \NA	\\ \hline
		\end{tabular}			}
		\label{benchmark_Comparion}			
	\end{table}
\end{landscape}

\section{Discussion and Future Work}
In this paper, we introduced \textsf{AMYTISS} as a software tool for parallel automated controller synthesis of large-scale discrete-time stochastic control systems. This tool is developed in C++/OpenCL for constructing finite MDPs and synthesizing controllers satisfying some high-level specifications. 
The tool can run in HPC platforms together with cloud computing services to reduce the problem of state-explosion. We proposed parallel algorithms to target HPC platforms and then implemented them within \textsf{AMYTISS}. 
As illustrated, \textsf{AMYTISS} significantly outperforms \textsf{FAUST}$^{\mathsf 2}$ and \textsf{StocHy} w.r.t. the computation time and memory usage. Providing a tool for large-scale \emph{continuous-time} stochastic control systems is under investigation as a future work.

\section{Acknowledgment}
The authors would like to thank Thomas Gabler for his help in implementing traditional serial algorithms for the purpose of analysis and then comparing with the parallel ones.

\bibliographystyle{alpha}
\bibliography{biblio}

\newcommand{\etalchar}[1]{$^{#1}$}
\begin{thebibliography}{WZK{\etalchar{+}}15}

\bibitem[ABC{\etalchar{+}}18]{abate2018arch}
A.~Abate, H.~Blom, N.~Cauchi, S.~Haesaert, A.~Hartmanns, K.~Lesser, M.~Oishi,
  V.~Sivaramakrishnan, S.~Soudjani, C.~I. Vasile, et~al.
\newblock {ARCH-COMP}18 category report: Stochastic modelling.
\newblock In {\em ARCH@ ADHS}, pages 71--103, 2018.

\bibitem[ABC{\etalchar{+}}19]{abate2019arch}
A.~Abate, H.~Blom, N.~Cauchi, K.~Degiorgio, M.~Fr{\"a}nzle, E.~M. Hahn,
  S.~Haesaert, H.~Ma, M.~Oishi, C.~Pilch, et~al.
\newblock {ARCH-COMP}19 category report: Stochastic modelling.
\newblock {\em EPiC Series in Computing}, 61:62--102, 2019.

\bibitem[Alt19]{Althof18}
M.~Althof.
\newblock Commonroad: Vehicle models (version 2018a). {T}ech. rep.
\newblock In {\em Technical University of Munich, 85748 Garching, Germany
  (October 2018)}, https://commonroad.in.tum.de. 2019.

\bibitem[BK08]{baier2008principles}
C.~Baier and J.-P. Katoen.
\newblock {\em Principles of model checking}.
\newblock MIT press, 2008.

\bibitem[BS96]{Bertsekas1996}
D.~P. Bertsekas and S.~E. Shreve.
\newblock {\em Stochastic Optimal Control: The Discrete-Time Case}.
\newblock Athena Scientific, 1996.

\bibitem[CA19]{StocHy19}
N.~Cauchi and A.~Abate.
\newblock \textsf{StocHy}: Automated verification and synthesis of stochastic
  processes.
\newblock In {\em TACAS'19}, volume 11428 of {\em Lecture Notes in Computer
  Science}, pages 247--264. 2019.

\bibitem[HH14]{hartmanns2014modest}
A.~Hartmanns and H.~Hermanns.
\newblock The modest toolset: An integrated environment for quantitative
  modelling and verification.
\newblock In {\em International Conference on Tools and Algorithms for the
  Construction and Analysis of Systems}, pages 593--598, 2014.

\bibitem[HS18]{HS_TAC19}
Sofie Haesaert and Sadegh Soudjani.
\newblock Robust dynamic programming for temporal logic control of stochastic
  systems.
\newblock {\em CoRR}, abs/1811.11445, 2018.

\bibitem[Jaj92]{jaja1992}
Joseph Jaja.
\newblock {\em An introduction to parallel algorithms}.
\newblock Addison-Wesley, 1992.

\bibitem[Kal97]{kallenberg1997foundations}
O.~Kallenberg.
\newblock {\em Foundations of modern probability}.
\newblock Springer-Verlag, New York, 1997.

\bibitem[KDS{\etalchar{+}}11]{kamgarpour2011discrete}
M.~Kamgarpour, J.~Ding, S.~Summers, A.~Abate, J.~Lygeros, and C.~Tomlin.
\newblock Discrete time stochastic hybrid dynamical games: Verification \&
  controller synthesis.
\newblock In {\em Proceedings of the 50th IEEE Conference on Decision and
  Control and European Control Conference}, pages 6122--6127, 2011.

\bibitem[KNP02]{kwiatkowska2002prism}
M.~Kwiatkowska, G.~Norman, and D.~Parker.
\newblock {PRISM}: Probabilistic symbolic model checker.
\newblock In {\em Proceedings of the International Conference on Modelling
  Techniques and Tools for Computer Performance Evaluation}, pages 200--204,
  2002.

\bibitem[KZ19]{pFaces}
Mahmoud Khaled and Majid Zamani.
\newblock p{F}aces: An acceleration ecosystem for symbolic control.
\newblock In {\em Proceedings of the 22nd ACM International Conference on
  Hybrid Systems: Computation and Control}, pages 252--257, 2019.

\bibitem[Lav19]{lavaei2019Thesis}
A.~Lavaei.
\newblock {\em Automated Verification and Control of Large-Scale Stochastic
  Cyber-Physical Systems: Compositional Techniques}.
\newblock PhD thesis, Technische Universit{\"a}t M{\"u}nchen, Germany, 2019.

\bibitem[LCGG13]{le2013mode}
E.l Le~Corronc, A.~Girard, and G.~Goessler.
\newblock Mode sequences as symbolic states in abstractions of incrementally
  stable switched systems.
\newblock In {\em Proceedings of the 52th IEEE Conference on Decision and
  Control}, pages 3225--3230, 2013.

\bibitem[LSMZ17]{lavaei2017compositional}
A.~Lavaei, S.~{Soudjani}, R.~Majumdar, and M.~Zamani.
\newblock Compositional abstractions of interconnected discrete-time stochastic
  control systems.
\newblock In {\em Proceedings of the 56th IEEE Conference on Decision and
  Control}, pages 3551--3556, 2017.

\bibitem[LSZ18a]{lavaei2018ADHS}
A.~Lavaei, S.~{Soudjani}, and M.~Zamani.
\newblock Compositional synthesis of finite abstractions for continuous-space
  stochastic control systems: A small-gain approach.
\newblock In {\em Proceedings of the 6th IFAC Conference on Analysis and Design
  of Hybrid Systems}, volume~51, pages 265--270, 2018.

\bibitem[LSZ18b]{lavaei2017HSCC}
A.~Lavaei, S.~Soudjani, and M.~Zamani.
\newblock From dissipativity theory to compositional construction of finite
  {M}arkov decision processes.
\newblock In {\em Proceedings of the 21st ACM International Conference on
  Hybrid Systems: Computation and Control}, pages 21--30, 2018.

\bibitem[LSZ19a]{lavaei2019NAHSJ}
A.~{Lavaei}, S.~{Soudjani}, and M.~Zamani.
\newblock Compositional abstraction-based synthesis of general {MDPs} via
  approximate probabilistic relations.
\newblock {\em arXiv: 1906.02930}, 2019.

\bibitem[LSZ19b]{lavaei2018CDCJ}
A.~Lavaei, S.~{Soudjani}, and M.~Zamani.
\newblock Compositional construction of infinite abstractions for networks of
  stochastic control systems.
\newblock {\em Automatica}, 107:125--137, 2019.

\bibitem[LSZ19c]{lavaei2019ECC}
A.~Lavaei, S.~{Soudjani}, and M.~Zamani.
\newblock Compositional synthesis of not necessarily stabilizable stochastic
  systems via finite abstractions.
\newblock In {\em Proceedings of the 18th European Control Conference}, pages
  2802--2807, 2019.

\bibitem[LSZ20a]{lavaei2019HSCC_J}
A.~Lavaei, S.~{Soudjani}, and M.~{Zamani}.
\newblock Compositional abstraction-based synthesis for networks of stochastic
  switched systems.
\newblock {\em Automatica}, 114, 2020.

\bibitem[LSZ20b]{lavaei2019NAHS}
A.~{Lavaei}, S.~{Soudjani}, and M.~{Zamani}.
\newblock Compositional abstraction of large-scale stochastic systems: A
  relaxed dissipativity approach.
\newblock {\em Nonlinear Analysis: Hybrid Systems}, 36, 2020.

\bibitem[LSZ20c]{lavaei2018ADHSJJ}
A.~Lavaei, S.~{Soudjani}, and M.~{Zamani}.
\newblock Compositional (in)finite abstractions for large-scale interconnected
  stochastic systems.
\newblock {\em IEEE Transactions on Automatic Control, DOI:
  10.1109/TAC.2020.2975812}, 2020.

\bibitem[LTS05]{li2005estimation}
W.~Li, E.~Todorov, and R.~E. Skelton.
\newblock Estimation and control of systems with multiplicative noise via
  linear matrix inequalities.
\newblock In {\em Proceedings of the American Control Conference}, pages
  1811--1816, 2005.

\bibitem[LZ19a]{lavaei2019LSS}
A.~Lavaei and M.~Zamani.
\newblock Compositional construction of finite {MDP}s for large-scale
  stochastic switched systems: A dissipativity approach.
\newblock {\em Proceedings of the 15th IFAC Symposium on Large Scale Complex
  Systems: Theory and Applications}, 52(3):31--36, 2019.

\bibitem[LZ19b]{lavaei2019CDC}
A.~Lavaei and M.~Zamani.
\newblock Compositional verification of large-scale stochastic systems via
  relaxed small-gain conditions.
\newblock In {\em Proceedings of the 58th IEEE Conference on Decision and
  Control}, pages 2574--2579, 2019.

\bibitem[MSSM19]{MSSM19}
K.~{Mallik}, A.~{Schmuck}, S.~{Soudjani}, and R.~{Majumdar}.
\newblock Compositional synthesis of finite-state abstractions.
\newblock {\em IEEE Transactions on Automatic Control}, 64(6):2629--2636, 2019.

\bibitem[Pnu77]{pnueli1977temporal}
A.~Pnueli.
\newblock The temporal logic of programs.
\newblock In {\em Proceedings of the 18th Annual Symposium on Foundations of
  Computer Science}, pages 46--57, 1977.

\bibitem[RWR16]{reissig2016feedback}
G.~Reissig, A.~Weber, and M.~Rungger.
\newblock Feedback refinement relations for the synthesis of symbolic
  controllers.
\newblock {\em IEEE Transactions on Automatic Control}, 62(4):1781--1796, 2016.

\bibitem[SA13]{esmaeil2013adaptive}
S.~Soudjani and A.~Abate.
\newblock Adaptive and sequential gridding procedures for the abstraction and
  verification of stochastic processes.
\newblock {\em SIAM Journal on Applied Dynamical Systems}, 12(2):921--956,
  2013.

\bibitem[SAM15]{SAM15}
S.~{Soudjani}, A.~Abate, and R.~Majumdar.
\newblock Dynamic {B}ayesian networks as formal abstractions of structured
  stochastic processes.
\newblock In {\em Proceedings of the 26th International Conference on
  Concurrency Theory}, pages 1--14, 2015.

\bibitem[SGA15]{FAUST15}
S.~{Soudjani}, C.~Gevaerts, and A.~Abate.
\newblock \textsf{FAUST}$^{\textsf{2}}$: Formal abstractions of
  uncountable-state stochastic processes.
\newblock In {\em TACAS'15}, volume 9035 of {\em Lecture Notes in Computer
  Science}, pages 272--286. 2015.

\bibitem[Sou14]{SSoudjani}
S.~Soudjani.
\newblock {\em Formal Abstractions for Automated Verification and Synthesis of
  Stochastic Systems}.
\newblock PhD thesis, Technische Universiteit Delft, The Netherlands, 2014.

\bibitem[SZ15]{shmarov2015probreach}
F.~Shmarov and P.~Zuliani.
\newblock \textsf{{ProbReach}}: verified probabilistic delta-reachability for
  stochastic hybrid systems.
\newblock In {\em Proceedings of the 18th International Conference on Hybrid
  Systems: Computation and Control}, pages 134--139, 2015.

\bibitem[VGO19]{vinod2019sreachtools}
A.~P. Vinod, J.~D. Gleason, and M.~M. Oishi.
\newblock \textsf{SReachTools}: {A} {MATLAB} stochastic reachability toolbox.
\newblock In {\em Proceedings of the 22nd ACM International Conference on
  Hybrid Systems: Computation and Control}, pages 33--38, 2019.

\bibitem[WZK{\etalchar{+}}15]{wang2015sreach}
Q.~Wang, P.~Zuliani, S.~Kong, S.~Gao, and E.~M. Clarke.
\newblock \textsf{{SReach}}: A probabilistic bounded delta-reachability
  analyzer for stochastic hybrid systems.
\newblock In {\em Proceedings of the International Conference on Computational
  Methods in Systems Biology}, pages 15--27, 2015.

\end{thebibliography}

\vspace{1cm}
{\small $^*$Both authors have contributed equally.}

\end{document}